\documentclass[12pt, draftclsnofoot, onecolumn]{IEEEtran}
\usepackage{hasan}
\usepackage{authblk}
\usepackage{xcolor}
\newcommand{\eff}{\tr{eff}}
\newcommand{\opt}{\tr{opt}}

\newcommand{\snr}{\texttt{SNR}}
\newcommand{\snreff}{\snr_{\tr{eff}}}
\newcommand{\snreffopt}{\snr_{\tr{eff},\tr{opt}}}
\newcommand{\codlat}{{\Lambda_{\tr{c}}}} % coding lattice
\newcommand{\shaplat}{\Lambda_{\tr{s}}} % shaping lattice
\newcommand{\Pout}{P_\tr{out}}
\newcommand{\Popt}{\mbf P_\tr{opt}}
\newcommand{\Ptopt}{\tilde{\mbf P}_\tr{opt}}
\newcommand{\Pin}{\mbf P_\tr{i}}
\newcommand{\vin}{\mbf v_\tr{i}}
\newcommand{\Pe}{P_\tr{e}}
\newcommand{\Pem}{P_{\tr{e},m}}
\newcommand{\Rt}{R_{\tr{t}}}

\newcommand{\Rif}{R_{\tr{IF}}}
\newcommand{\Jt}{\tilde{J}}
\newcommand{\Jmu}{\hat{J}(\mu{})}
\newcommand{\Pmu}{\mbf{P}(\mu{})}
\newcommand{\Om}{O (M)}
\newcommand{\om}{\mfr o (M)}
\newcommand{\Pl}{\mbf P_{\ell}}
\newcommand{\Pln}{\mbf P_{\ell+1}}
\newcommand{\Rl}{\mbf R_{\ell}}
\newcommand{\Sl}{\mbf S_{\ell}}
\newcommand{\mul}{\mu_\ell}
\newcommand{\mdort}{\lambda_1^{(\tr o)}}
\newcommand{\mduni}{\lambda_1^{(\tr u)}}
\newcommand{\Linv}{\mbf L^{-1}}
\newcommand{\Lp}{\mbf L_{\tr P}}
\newcommand{\Lpt}{\mbf L_{\breve{\tr P}}}

\newcommand{\lat}[1]{\Lambda(#1)}
\newcommand{\dualat}[1]{\Lambda^*(#1)}
\newcommand{\sucmin}[3]{\lambda_{#1}^{#3}(#2)}

\tikzset{
  curve1/.style={TolDarkBlue, thick, mark=*, mark size=1.6pt},
  curve2/.style={TolLightBrown, thick, mark=square*, mark size=1.3pt},
  curve3/.style={TolLightGreen, thick, mark=triangle*, mark size=1.5pt},
  curve4/.style={TolDarkBrown, thick, mark=diamond*, mark size=1.5pt},
}

\IEEEaftertitletext{\vspace{-1.8\baselineskip}}

\makeatletter
\newcommand{\removelatexerror}{\let\@latex@error\@gobble}
\makeatother
\IEEEoverridecommandlockouts

\title{Steepest Gradient-Based Orthogonal Precoder For Integer-Forcing MIMO}
\author{Mohammad Nur Hasan, Brian M. Kurkoski, Amin Sakzad, and Emanuele Viterbo
  \thanks{Mohammad Nur Hasan and Brian M. Kurkoski are with School of Information Science, Japan Advanced Institute of Science and Technology (JAIST), 1-1 Asahidai, Nomi, Ishikawa, Japan (e-mail: hasan-mn@ieee.org, kurkoski@ieee.org).}
  \thanks{Amin Sakzad is with Faculty of Information Technology, Monash University, Clayton, Melbourne, Vic. 3800, Australia (e-mail: amin.sakzad@monash.edu).}
  \thanks{Emanuele Viterbo is with Department of Electrical and Computer Systems Engineering, Monash University, Clayton, Melbourne, Vic. 3800, Australia (e-mail: emanuele.viterbo@monash.edu).}
}
\begin{document}
%\sloppy
\maketitle

\begin{abstract}
In this paper, we develop an orthogonal precoding scheme for integer-forcing (IF) linear receivers using the steepest gradient algorithm. Although this scheme can be viewed as a special case of the unitary precoded integer-forcing (UPIF), it has two major advantages. First, the orthogonal precoding outperforms its unitary counterpart in terms of achievable rate, outage probability, and error rate. We verify this advantage via theoretical and numerical analyses. Second, it exhibits lower complexity as the dimension of orthogonal matrices is half that of unitary matrices in the real-valued domain. For finding ``good'' orthogonal precoder matrices, we propose an efficient algorithm based on the steepest gradient algorithm that exploits the geometrical properties of orthogonal matrices as a Lie group. The proposed algorithm has low complexity and can be easily applied to an arbitrary MIMO configuration. We also confirm numerically that the proposed orthogonal precoding outperforms UPIF type II in some scenarios and the X-precoder in high-order QAM schemes, e.g., $64$- and $256$-QAM.
\end{abstract}

\section{Introduction}
\label{sec:introduction}
\IEEEPARstart{F}{uture} wireless networks are facing unprecedented challenges as the number of wirelessly-connected devices such as smartphones, tablets, computers, and sensors is dramatically increasing. Furthermore, the emergence of abundant software applications demanding high quality media, e.g., images and videos, results in the tremendous increase of the global network traffic. This situation leads to the demands of massive wireless network access and high data transmission rate. The scarcity of the available spectrum frequency makes these challenges more difficult to overcome. The use of multiple antennas at both  transmitter and receiver in a wireless communication system known as the multiple-input multiple-output (MIMO) system \cite{Telatar99capacityof} has emerged as one key technology to cope with the above problems. Exploiting multi-path scattering, MIMO offers significant improvement in terms of transmission reliability (diversity gain) and data transmission rate.

To realize the advantages of MIMO, it is important to design an optimal or near-optimal receiver. A maximum likelihood (ML) receiver has optimal rates and probability of error \cite{ViterboB99SD}. However, its complexity increases exponentially with respect to the number of antennas. As alternatives, zero-forcing (ZF) or minimum mean square error (MMSE) receivers are often employed \cite{Tse05fundamentalsof}. These receivers apply a linear transformation such that the MIMO channel can be seen as a sequence of single-input single-output (SISO) channels, and hence, the decoding complexity is greatly reduced. However, this advantage comes with the cost of a performance loss which can be significant especially in the low signal-to-noise power ratio (SNR) regime.  Zhan \textit{et al.} proposed a MIMO linear receiver called integer-forcing (IF) receiver \cite{zhan14IF} which achieves significantly better error performance than ZF and MMSE receivers with nearly the same decoding complexity  for slow-fading channels. In the IF receiver framework, the transmitter employs nested lattice codes and the receiver approximates the channels with a ``good'' full rank \textit{integer} matrix $\mbf A$. Since an integer linear combination of lattice codewords is again a codeword, the receiver can use SISO decoding to decode each linear combination, and subsequently recover the transmitted messages by solving a simple linear equation system. It has been shown that IF receivers achieve the optimal diversity-multiplexing tradeoff (DMT) \cite{ZhengT03,OrdentlichE15} and yield numerical error performance that is quite close to that of the optimal ML receiver \cite{zhan14IF,SakzadHV13}.

While the advantages of MIMO can be achieved  when the channel state information (CSI) is only available at the receiver, these  can be further enhanced when the transmitter has some level of knowledge of CSI. The transmitter exploits CSI for encoding information symbols prior to transmissions to increase the reliability against the channel fluctuations; this technique is known as \textit{precoding} \cite{VuP07MimoPrecoding}. Many precoding schemes are designed for MIMO with quadrature amplitude modulations (QAM) and ML receivers. For instance, Vrigneau \textit{et al.} \cite{Vrigneau08} proposed a specific precoding scheme for $4$-QAM MIMO systems with ML receivers. This precoding is optimal and has been shown to outperform all MMSE receiver-based precodings. However, despite its optimality, it is hard to further extend the idea to higher-order QAM because of its high complexity. In \cite{XYcodes}, Mohammed \textit{et al.} proposed precoding schemes for more general QAM with ML receivers, namely X- and Y-precoders. These precoding schemes can achieve error performance close to that of \cite{Vrigneau08} and can be easily employed for an arbitrary MIMO configuration. However, when full transmission rate is used, X- and Y-precoders cannot achieve full diversity gain. Moreover, since they are designed based on the minimum distance of the received QAM constellations, the error performance degrades as the constellation size increases.

The subject of this paper is precoding schemes for MIMO with integer-forcing receivers (IF-MIMO). The performance of this kind of precoding is not dictated by the minimum distance of received constellations, and hence, it can excel in high-order modulation schemes. In \cite{SakzadV15UnitaryPrecoding}, Sakzad and Viterbo proposed \textit{unitary precoded integer-forcing} (UPIF), a precoding scheme designed for IF-MIMO where the precoder matrices are from groups of unitary matrices. They showed that UPIF achieves full diversity gain while allowing full rate transmission. Two types of UPIF were introduced. The first type of precocder (UPIF~I) is designed for each channel realization based on the \textit{minimum distance} of a lattice generated by the precoder matrix. The second type of precoder (UPIF~II) is designed for all channel realizations based on the \textit{minimum product distance} \cite{FluckigerOV04} of the generated lattice. In this paper we are particularly interested in UPIF~I where the precoder matrix adapts to each channel realization. Finding the optimal precoder matrix of UPIF~I is a hard problem due to the involvement of the unitary constraint \cite{EdelmanOrtho99} and the lattice minimum distance problem \cite{FinckeP85 ,Bremner11,SchnorrE94}. For $2 \times 2$ MIMO systems, a simple parameterization technique finds the optimal UPIF~I precoder matrix \cite{SakzadV15UnitaryPrecoding}. But for higher-order MIMO, this technique is computationally expensive because an exhaustive search over multiple parameters is required. 

This paper addresses this problem and proposes an efficient algorithm for finding good orthogonal precoders matrices that are applicable to any MIMO dimension. The summary and  contributions  of this work are as follows.
\begin{enumerate}
    \item In \cite{SakzadV15UnitaryPrecoding} it is shown that the search space for optimal UPIF~I precoder matrices is groups of unitary matrices. However, in this paper we argue that it is sufficient and even superior to only search over groups of orthogonal matrices.\footnote{Groups of orthogonal matrices are sub-groups of groups of unitary matrices.} Unitary precoder matrices do not guarantee better achievable rate and outage probability than orthogonal precoder matrices; this is shown using Propositions~\ref{proposition:snr-bound-orthogonal}~and~\ref{proposition:snr-bound-unitary}.  Via numerical evaluations we confirm that indeed the orthogonal precoder outperforms its unitary counterpart in terms of achievable rate, outage probability, and error rate. Besides the performance advantage, the orthogonal precoder also has lower complexity as the dimension of orthogonal matrices is half that of unitary matrices in real-valued domain. In other words, we show that the orthogonal precoder is more favorable in terms of both performance and complexity compared to unitary precoders for UPIF~I.

  \item We propose an efficient algorithm for finding good orthogonal precoder matrices. This algorithm is based on the steepest gradient algorithm and exploits the geometrical properties of orthogonal matrices as a Lie group \cite{Plumbley05,EdelmanOrtho99,AbrudanEK08}. The main difficulty of the optimization problem comes from the simultaneous inclusions of (i) an orthogonality constraint and (ii) the lattice minimum distance problem. Without the minimum distance problem, we could immediately use existing steepest gradient algorithms. However, the inclusion of (ii) makes the optimization problem non-differentiable and much harder. Our approach is to divide the problem into two sub-problems, and develop algorithms based on  steepest gradient and random search algorithms to solve them. Discussion of the proposed algorithm is presented in Section~\ref{sec:find-prec-matr}. Compared to the parameterization technique \cite{SakzadV15UnitaryPrecoding,Raffanetti70}, the proposed algorithm has lower complexity --- the proposed algorithm has polynomial complexity of $\mcal O(M^4 \log M)$, while the parameterization technique has exponential $\mcal O (\nu^{M(M-1)/2)} M^4 \log M)$, where $M$ is the number of antennas and $\nu$ is a constant, cf. Section~\ref{sec:complexity}.
    
  \item We present and analyze the results of computer simulations comparing the proposed schemes with existing schemes. The numerical results show that:
  \begin{itemize}
  \item Orthogonal precoder matrices are superior to unitary precoder matrices for integer-forcing MIMO. 
  \item Despite its lower complexity, the proposed steepest gradient-based algorithm achieves performance identical to the parameterization technique.
  \item Even though X-precoders are designed specifically for QAM, our proposed schemes are remarkably better (in terms probability of error) in high-order QAM schemes, e.g., $64$- and $256$-QAM.
  \item The proposed schemes outperform UPIF~II in some scenarios, e.g., $4 \times 4$ MIMO.    
  \end{itemize}
\end{enumerate}  
Compared to our earlier conference paper \cite{HasanISIT19}, this paper provides Propositions~\ref{proposition:error-prob-bound} and \ref{proposition:snr-bound-unitary}, their proofs, and detailed performance analyses. This paper also presents details of computational complexity analysis in Section~\ref{sec:complexity} and adds substantial numerical results to validate the advantages of the proposed schemes. 

\textit{Notation:} Let $\mbb R, \mbb C, \mbb Z$ be the real, complex, and integer numbers, respectively. $\mbb Z[i]$ denotes the Gaussian integers. For any complex number, $\Im(\cdot)$ and $\Re(\cdot)$ denote its real and imaginary components, respectively. Let $O(n)$ and $U(n)$ respectively denote the orthogonal and unitary  groups of dimension $n$.\footnote{Orthogonal and unitary groups are groups of orthogonal and unitary matrices, respectively.} Boldface lowercase letters denote vectors, e.g., $\mbf a \in \mbb Z^n$, while boldface uppercase letters denote matrices, e.g., $\mbf A \in \mbb Z^{n \times n}$. The Hermitian and the regular transpose operations are expressed by $(\cdot)^H$ and $(\cdot)^T$, e.g., $\mbf A^H$ and $\mbf A^T$, respectively. The inversion of the regular transpose is denoted by $(\cdot)^{-T}$, e.g., $\mbf A^{-T} \triangleq (\mbf A^T)^{-1}$. The matrix exponential is defined as $\exp(\mbf A) \triangleq \sum_{m=0}^{\infty}\frac{\mbf A^m}{m!}$. The general logarithm is with base $2$, unless otherwise stated. 

\section{Preliminaries}
\label{sec:preliminaries}
In this section we recall some essential lattice-related definitions that are useful for understanding our proposed technique. A lattice is a discrete subgroup of the Euclidean space  with vector addition operation. Formally, lattices are defined as follows.
\begin{mydefinition}[Real-valued lattice]\label{def:real-lattice}
  Given a full-rank generator matrix $\mbf G \in \mbb R^{n \times n}$, the real-valued lattice $\lat{\mbf G}$ is composed of all integral combinations of the column vectors of $\mbf G$, i.e.,\footnote{We use only $\Lambda$ to denote a lattice when its generator matrix is undefined.}
  \begin{align}
    \lat{\mbf G} = \{\mbf G \mbf a : \mbf a \in \mbb Z^n\}. \label{eq:real-lattice}
  \end{align}
\end{mydefinition}
\begin{mydefinition}[Dual lattice]\label{def:dual-lattice}
  For a real-valued lattice $\lat{\mbf G}$ with a full-rank generator matrix $\mbf G \in \mbb R^{n \times n}$, the dual lattice is 
  \begin{align}
    \dualat{\mbf G} &\triangleq \lat{\mbf G^{-T}}\\
                    &= \{\mbf G^{-T}\mbf a:\mbf a \in \mbb Z^n\}. \label{eq:dual-lattice}
  \end{align}
\end{mydefinition}

\begin{mydefinition}[Complex-valued lattice]\label{def:complex-lattice}
  Given a full-rank generator matrix $\breve{\mbf G} \in \mbb C^{n \times n}$, the complex-valued lattice $\lat{\breve{\mbf G}}$ is defined similarly to the real-valued lattice as
  \begin{align}
    \lat{\breve{\mbf G}} = \{\breve{\mbf G} \breve{\mbf a} : \breve{\mbf a} \in \mbb Z[i]^n\}.
  \end{align}
\end{mydefinition}

In the following, a few important notions associated with lattices is given.
\begin{mydefinition}[Successive minima]\label{def:successive-minima}
  For an $n$-dimensional lattice $\lat{ \mbf G}$ generated by a full-rank matrix $\mbf G$, the $l$-th successive minimum, $1 \leq l \leq n$, is defined as
  \begin{align}
    \sucmin{l}{\mbf G}{} \triangleq \min_{\mbf v_1,...,\mbf v_l \in \lat{\mbf G}} \max \{\norm{\mbf v_1}, ..., \norm{\mbf v_l}\}, \label{eq:successive-minima}
  \end{align}
  where the minimum is taken over all sets of $l$ linearly independent vectors in $\lat{\mbf G}$. In other words, $\sucmin{l}{\mbf G}{}$ is the smallest real number $r$ such that there exist $l$ linearly independent vectors $\mbf v_1, ..., \mbf v_l \in \lat{\mbf G}$ with $\norm{\mbf v_1}, ..., \norm{\mbf v_l} \leq r$. Note that the first successive minimum of $\lat{\mbf G}$, i.e., $\sucmin{1}{\mbf G}{}$, is its minimum distance. The successive minima are non-decreasing,
\begin{align}
  \sucmin{1}{\mbf G}{} \leq \sucmin{2}{\mbf G}{} \leq \cdots \leq \sucmin{n}{\mbf G}{}.\label{eq:nondec-suc-min}
\end{align}
\end{mydefinition}
\begin{mydefinition}[Fundamental Voronoi region]\label{def:voronoi-region}
  The fundamental Voronoi region of an $n$-dimensional real-valued lattice $\Lambda$, denoted by $\mcal V_{\Lambda}$, consists of all points of the underlying space that are closer to the origin $\mbf 0$ than any other lattice point, i.e.,
  \begin{align}
    \mcal V_{\Lambda} = \left\{ \mbf r \in \mbb R^n : \abs{\mbf r} \leq \abs{\mbf r - \mbf t} \tr{for all } \mbf t \in \Lambda \backslash \mbf 0 \right\}. 
  \end{align}
  The Voronoi region associated with each $\mbf t \in \Lambda$ is a shift of $\mcal V_{\Lambda}$ by $\mbf t$. The fundamental Voronoi region of a complex-valued lattice is defined similarly.
\end{mydefinition}
\begin{mydefinition}[Nested lattice code \cite{ErezZ04Awgn,ZamirSE02,Kurkoski16}]\label{def:nested-lattice}
  Given two lattices $\codlat$ and $\shaplat$ where $\shaplat \subset \codlat$, the nested lattice code $\mcal C$ is defined as the coset leaders of the quotient group $\codlat/\shaplat$ that are within the fundamental Voronoi region of $\shaplat$, i.e.,
  \begin{align}
    \mcal C = \codlat \cap \mcal V_{\shaplat}.
  \end{align}
  $\codlat$ is the fine lattice used for coding and $\shaplat$ is the coarse lattice used for shaping. The rate of $\mcal C$ is
\begin{align}
  R = \frac 1 n \log \abs{\mcal C}.
\end{align}
\end{mydefinition}
\section{IF MIMO with Orthogonal Precoder}
\label{sec:if-mimo-ortho-precoder}

\subsection{System Model}
\label{sec:system-model}
Without loss of generality, we consider a point-to-point MIMO system where each transmission end is equipped with $M$ antennas, i.e., an $M \times M$ MIMO system. The channels are assumed to be quasi-static flat-fading, remaining constant over one coherence interval. CSI is known to both transmitter and receiver. Denoted by $\mbf H \in \mbb C^{M \times M}$, the channel matrix is decomposed to $\mbf H = \mbf W \mbf D \mbf V^H$ using the singular value decomposition (SVD). $\mbf W, \mbf V \in \mbb C^{M \times M}$ are unitary matrices, i.e., $\mbf{WW}^H = \mbf{VV}^H = \mbf I$, and $\mbf D \triangleq \tr{diag}(d_1,d_2,...,d_M) \in \mbb R^{M \times M}$ is a diagonal matrix with $d_1\geq d_2\geq \cdots \geq d_M$.

Let $\mcal C$ be a codebook of a nested lattice $\codlat/\shaplat \subset \mbb C^n$ with coding rate $R$. Let $\mbf w_m$, $m = 1,...,M$, be information messages to be transmitted across MIMO channels. These messages are encoded to lattice codewords $\mbf x_m \in \mcal C$ using a bijective mapping $\mcal E$, i.e., $\mcal E(\mbf w_m) = \mbf x_m$. Each $\mbf x_m$ satisfies $\frac{1}{n}\mbb E ||\mbf x_m||^2 = \gamma$. Let $\mbf X =[\mbf x_1 \; \cdots \; \mbf x_M]^T \in \mbb C^{M \times n}$. Prior to transmissions, $\mbf X$ is precoded such that $\mbf X_{\tr{prec}} = \mbf V \mbf P \mbf X$, where $\mbf P \in \mbb R^{M \times M}$ is an orthogonal matrix. We refer to the matrix $\mbf P$ as the \textit{precoder matrix}, which is subject to the optimization problem in this work. The received signal at the receiver is 
\begin{align}
  \mbf Y = \mbf H \mbf X_{\tr{prec}} + \mbf Z. \label{eq:mimo} 
\end{align} 
The entries of $\mbf H$ and $\mbf Z \in \mbb C^{N\times n}$ are i.i.d. complex Gaussian random variables  $\sim \mcal{CN}(0,1)$. We assume that random dithering is employed to ensure that the $\mbf x_m$ is uniformly distributed over the fundamental Voronoi region of $\shaplat$. However, for simplicity, we omit the dithering notations from the exposition. Upon receiving $\mbf Y$, the receiver multiplies it by $\mbf W^H$, and thus,
\begin{align}
  \tilde{\mbf{Y}} = \mbf W^H \mbf Y &= \mbf W^H \mbf H \mbf X_{\tr{prec}} + \mbf W^H \mbf Z \\
                                    &= \mbf W^H \mbf W \mbf D \mbf V^H \mbf V \mbf P \mbf X + \mbf W^H \mbf Z \\
                                    &= \mbf D \mbf P \mbf X + \tilde{\mbf Z}, \label{eq:mimo-ori}
\end{align}
with $\tilde{\mbf Z} = \mbf W^H \mbf Z$ whose entries still follow $\mcal{CN}(0,1)$ because $\mbf W$ is unitary.

The receiver employs an IF receiver \cite{zhan14IF} which transforms the resulting channel in \eqref{eq:mimo-ori} into $M$  effective point-to-point sub-channels. Hence the receiver can decode the transmitted messages using a SISO decoding rather than joint decoding across all receive antennas. In principle, the IF receiver approximates the resulting MIMO channel $\mbf D \mbf P$  with an invertible integer matrix\footnote{Note that if $\mbf P$ is a unitary matrix (complex-valued), then $\mbf A \in \mbb Z[i]^{M \times M}$ and $\mbf B \in \mbb C^{M \times M}$.} $\mbf A \in \mbb Z^{M \times M}$ by selecting an equalizing matrix $\mbf B \in \mbb R^{M \times M}$ and computes\footnote{$\bmod \; \shaplat$ is modulo operation on each row of the corresponding matrix with respect to the shaping lattice $\shaplat$.}
\begin{align}
  \mbf Y_{\eff} &= [\mbf B \tilde{\mbf Y}] \bmod  \Lambda_{\tr s} \\
          &= [\mbf B \mbf D \mbf P \mbf X + \mbf B \tilde{\mbf Z}] \bmod  \Lambda_{\tr s} \\ \label{eq:BFilter}
          &= [\mbf A \mbf X + (\mbf B \mbf D \mbf P - \mbf A) \mbf X + \mbf B \tilde{\mbf Z}] \bmod  \Lambda_{\tr s}. 
\end{align}

Let $\mbf y_{\eff, m}^T$, $\mbf a_m^T$, and $\mbf b_m^T$  be the $m$-th rows of $\mbf Y_{\eff}$, $\mbf A$, and $\mbf B$, respectively. The effective received signal at sub-channel $m$ can be written as
\begin{align}
  \mbf y_{\eff,m}^T &=  [\mbf a_m^T \mbf X + (\mbf b_m^T \mbf D \mbf P - \mbf a_m^T) \mbf X + \mbf b_m^T \tilde{\mbf Z}] \bmod \shaplat\\
                      &= [\mbf c_m^T + \mbf z_{\eff, m}^T] \bmod \shaplat, \label{eq:p2p-eff-channel}
\end{align}
where $\mbf c_m^T = \mbf a_m^T \mbf X \bmod \shaplat$ is the desired linear combination, and
\begin{align}
  \mbf z_{\eff, m}^T =  [(\mbf b_m^T \mbf D \mbf P - \mbf a_m^T) \mbf X + \mbf b_m^T \tilde{\mbf Z}] \bmod \shaplat   \label{eq:z-eff}
\end{align}
is the effective noise at sub-channel $m$.

Owing to the linearity property of $\mcal C$, the linear combination $\mbf c_m$ happens to be a codeword, and thus, the next step of the IF receiver is to decode $\mbf c_m$ from the effective point-to-point sub-channel in \eqref{eq:p2p-eff-channel}. Let $\hat{\mbf c}_m$ be the estimate of $\mbf c_m$. $\hat{\mbf c}_m$ is obtained using $\hat{\mbf c}_m = Q_\codlat(\mbf y_{\eff, m})$, where $Q_\codlat(\cdot)$ is the decoding or quantization function with respect to $\codlat$. Let $\hat{\mbf C} = [\hat{\mbf c}_1, ..., \hat{\mbf c}_M]^T$, and $\hat{\mbf X}$ and $\hat{\mbf w}_m$ be the estimates of $\mbf X$ and $\mbf w_m$, respectively. The transmitted symbols are obtained by solving $\hat{\mbf X} = \mbf A^{-1} \hat{\mbf C}$, and finally the information messages are recovered using $\hat{\mbf w}_m = \mcal E^{-1} (\hat{\mbf x}_m)$.

\subsection{Performance Metrics}
\label{sec:performance-metrics}
Consider the performance of this MIMO system. First, define the variance of $\mbf z_{\eff, m}$ as
\begin{align}
  \sigma_{\eff, m}^2 &\triangleq \frac 1 n \mbb E \norm{(\mbf b_m^T \mbf D \mbf P - \mbf a_m^T) \mbf X + \mbf b_m^T \tilde{\mbf Z}}^2 \nonumber \\
                         &=\gamma \norm{\mbf b_m^T \mbf D \mbf P - \mbf a_m^T}^2 + \big\lVert \mbf b_m^T \big\lVert^2. \label{eq:noise-eff-var}
\end{align}
To achieve a reliable communication system, $\mbf b_m$ should be chosen such that the effective noise variance $\sigma_{\eff, m}^2$ is minimized. The optimal $\mbf b_m$ is \cite{SakzadV15UnitaryPrecoding}
\begin{align}
  \mbf b_{\tr{opt}, m}^T = \gamma \mbf a_m^T (\mbf D \mbf P)^T(\mbf I + \gamma \mbf D \mbf P (\mbf D \mbf P)^T)^{-1}.
\end{align}
Substituting $\mbf b_{\tr{opt}, m}^T$ into \eqref{eq:noise-eff-var} results in
\begin{align}
  \sigma_{\eff, m}^2 &= \gamma \mbf a_m^T (\mbf I + \gamma  (\mbf D \mbf P)^T \mbf D \mbf P)^{-1} \mbf a_m \\
                        &= \gamma \mbf a_m^T \mbf P^T(\mbf I + \gamma  \mbf D^T \mbf D)^{-1}\mbf P \mbf a_m. 
\end{align}
Because $(\mbf I + \gamma  \mbf D^T \mbf D)^{-1}$ is a positive definite matrix, it admits Cholesky decomposition
\begin{align}
  (\mbf I + \gamma  \mbf D^T \mbf D)^{-1} = \mbf L \mbf L^T.
\end{align}
Now let
\begin{align}
\Lp \triangleq \mbf P^T \mbf L.\label{eq:Lp}
\end{align}
Hence, $\sigma_{\eff, m}^2$ can be expressed as
\begin{align}
  \sigma_{\eff, m}^2 &= \gamma \mbf a_m^T \mbf P^T \mbf L \mbf L^T \mbf P \mbf a_m \\
                     &= \gamma \norm{\Lp^T \mbf a_m}^2.
\end{align}

Define the effective SNR of the worst sub-channel, i.e., the channel with the highest effective noise variance, as
\begin{align}
  \snreff &\triangleq \min_{m = 1, ..., M} \frac {\frac 1 n \mbb E \norm{\mbf c_m}^2} {\sigma_{\eff, m}^2}\\
          % &= \min_{m = 1,...,M} \frac \gamma {\gamma\norm{\Lp^T \mbf a_m}^2} \\
          &= \min_{m = 1,...,M} \frac 1 {\norm{\Lp^T \mbf a_m}^2}.
\end{align}
Note that because $\mbf c_m$ is a codeword, $\frac 1 n \mbb E \norm{\mbf c_m}^2 = \gamma$. Clearly, to recover the information messages, all $\mbf c_m$'s must be decoded correctly. Therefore, the matrix $\mbf A$ has to be chosen such that $\snreff$ is maximized. Define the optimal matrix $\mbf A$ as
\begin{align}
  \mbf A_\opt &= \argmax_{\substack{\mbf A \in \mbb Z^{M \times M}\\ \tr{det}(\mbf A) \neq 0}} \min_{m = 1,...,M} \frac 1 {\norm{\Lp^T \mbf a_m}^2}\\
  &= \argmin_{\substack{\mbf A \in \mbb Z^{M \times M}\\ \tr{det}(\mbf A) \neq 0}} \max_{m = 1,...,M} \norm{\Lp^T \mbf a_m}^2 \label{eq:A-opt}.
\end{align}
If $\mbf A_\opt$ is employed, then we have the optimal $\snreff$ as
\begin{align}
  \snreffopt = \frac 1 {\sucmin{M}{\Lp^T}{2}}, \label{eq:snreff-opt}
\end{align}
where $\sucmin{M}{\Lp^T}{}$ is the largest successive minimum of the lattice $\lat{\Lp^T}$, see the definition of successive minima given in \eqref{eq:successive-minima}. Finding $\mbf A_\opt$ is one of crucial problems in the IF framework. Because this problem is equivalent to finding successive minima of a lattice, we can conveniently employ the sphere decoding algorithms \cite{ViterboB99SD,FinckeP85} or the LLL algorithms \cite{Lenstra82LLL,Gan09}. We can also use the recently proposed algorithms specifically for IF-MIMO \cite{SakzadHV13,LyuBoostedLLL,Liu16,DingSMPIF15}.

Assume that a ``good'' nested lattice code $\mcal C$ \cite{NazerG11,ErezZ04Awgn,ZamirSE02,zhan14IF} is employed at the transmitter. In the IF receiver framework, the worst sub-channel constitutes a performance bottleneck. Therefore, if the rate of $\mcal C$ satisfies
\begin{align}
  R < \log(\snreffopt),
\end{align}
then all sub-channels $m = 1,...,M$ can decode their linear combination $\mbf c_m$ with a low error probability. This implies that the achievable rate of this MIMO system is
\begin{align}
  \Rif = M \log(\snreffopt). \label{eq:achievable-rate}
\end{align}

Let $\Rt $ be the target rate of the system. The outage probability of the system is defined as
\begin{align}
  \Pout &\triangleq \Pr(\Rif < \Rt)\\
        &= \Pr(M \log(\snreffopt) < \Rt)\\
        &= \Pr\big(\snreffopt < 2^{\Rt/M}\big).\label{eq:outage-prob}
\end{align}

From \eqref{eq:achievable-rate} and \eqref{eq:outage-prob}, we know that to improve the performance in terms of achievable rate and outage probability, $\snreffopt$ should be maximized. This maximization is rather difficult because $\snreffopt$ is a function of the largest successive minimum of a lattice. However, we can bound $\snreffopt$ with the minimum distance of its dual lattice, which makes the optimization easier. For this purpose, we use the following proposition.
\begin{myproposition}\label{proposition:snr-bound-orthogonal}
  Consider the aforementioned IF-MIMO system with an orthogonal precoder matrix $\mbf P$. The effective SNR of the worst sub-channel is lower bounded by
  \begin{align}
    \snreffopt \geq \frac  {\sucmin{1}{\Lp^{-1}}{2}}{M^2}, \label{eq:snreff-bound} 
  \end{align}
  where $\Lp$ is defined in (\ref{eq:Lp}) and $\sucmin{1}{\Lp^{-1}}{}$ is the minimum distance of lattice $\lat{\Lp^{-1}}$, which is the dual lattice of $\lat{\Lp^T}$.
\end{myproposition}

\begin{proof}
  The proof is given in Appendix~\ref{sec:proof-proposition-ortho}.
\end{proof}

Using Proposition~\ref{proposition:snr-bound-orthogonal}, we now can bound the achievable rate of the system as
\begin{align}
  \Rif &= M \log(\snreffopt)\\
       &\geq M \log \Big(\frac{\sucmin{1}{\Lp^{-1}}{2}}{M^2} \Big)\\
       &= 2M \big(\log (\sucmin{1}{\Lp^{-1}}{} ) - \log(M) \big),\label{eq:rate-bound}
\end{align}
and the outage probability as
\begin{align}
  \Pout &= \Pr\big(\snreffopt < 2^{\Rt/M}\big)\\
        &\leq \Pr \Big( \frac  {\sucmin{1}{\Lp^{-1}}{2}}{M^2}  < 2^{\Rt/M} \Big)\\
        &= \Pr \Big(   \sucmin{1}{\Lp^{-1}}{2}  < M^22^{\Rt/M} \Big). \label{eq:outage-bound}
\end{align}

Define the error probability of the system as
\begin{align}\label{eq:error-prob}
  \Pe = \Pr\big((\hat{\mbf w}_1,...,\hat{\mbf w}_M) \neq (\mbf w_1,...,\mbf w_M)\big).
\end{align}
This error probability is dependent of the nested lattice code $\mcal C$ employed at the system. From a practical point of view we may consider $2^{2q}$-QAM constellations for a positive integer $q$, e.g., $4$-QAM, $16$-QAM, and $64$-QAM. These constellations are equivalent to the nested lattice code $\codlat / \shaplat$ with $\codlat = \alpha \mbb Z[i]$ and $\shaplat = 2^q \codlat$, where $\alpha$ is a positive real number. Employing this code, the error probability of the system is given by the following proposition.

\begin{myproposition}\label{proposition:error-prob-bound}
  If nested lattice code $\codlat / \shaplat$, with $\codlat = \alpha \mbb Z[i]$ and $\shaplat = 2^q \codlat$, where $1<q \in \mbb Z$ and $\alpha = \sqrt{6\gamma}/2^{2q}$, is employed in an $M \times M$ IF-MIMO system, the error probability is bounded as
  \begin{align}
    \Pe &\leq 4M \exp\Big( -\frac {3\sucmin{1}{\Lp^{-1}}{2}} {2^{4q+1}M^2}\Big),\label{eq:error-bound}
  \end{align}
  where $\Lp$ is defined in (\ref{eq:Lp}).
\end{myproposition}
\begin{proof}
  See Appendix~\ref{sec:proof-proposition-error-bound}.
\end{proof}

\subsection{Problem Statement}
\label{sec:problem-statement}
The performance metrics derived in \eqref{eq:rate-bound}, \eqref{eq:outage-bound}, and \eqref{eq:error-bound} suggest that to achieve a good performance in terms of achievable rate, outage probability, and error probability, we should choose precoder matrix $\mbf P$ such that $\sucmin{1}{\Lp^{-1}}{2}$ is maximized. Formally, we define the problem of finding the optimal $\mbf P$ as
\begin{align}
  \mbf \Popt &= \argmax_{\mbf P \in \Om} \sucmin{1}{\Lp^{-1}}{2} \label{eq:main-optim-problem-1}\\
             &= \argmax_{\mbf P \in \Om} \min_{\mbf v \in \mbb Z^M  \backslash \mbf 0} \norm{\Linv \mbf P \mbf v}^2.\label{eq:main-optim-problem}
\end{align}
In other words, we have to find an orthogonal matrix $\mbf P$ such that the minimum distance of lattice $\lat{\Linv \mbf P}$ is maximized.

Based on \eqref{eq:main-optim-problem}, one may argue that unitary precoder matrices can yield a larger $\sucmin{1}{\Lp^{-1}}{}$ than the orthogonal one. Indeed, that is the case. But, recall that we derive the bounds on performance metrics in \eqref{eq:rate-bound}, \eqref{eq:outage-bound}, and \eqref{eq:error-bound} in order to ease the optimization process. The performance of the system is more directly affected by $\snreffopt$ or $\sucmin{M}{\Lp^{T}}{}$ rather than by $\sucmin{1}{\Lp^{-1}}{}$. We introduce the following proposition for the case where a unitary matrix is employed as the precoder matrix.
\begin{myproposition}\label{proposition:snr-bound-unitary}
  Consider a precoded IF-MIMO system similar to  the aforementioned one except that the precoder matrix is unitary. Let $\breve{\mbf P} \in U (M)$ be the precoder matrix and $\Lpt = \breve{\mbf P}^H \mbf L$ be the matrix corresponding to \eqref{eq:Lp} in the orthogonal precoder case. The effective SNR of the worst sub-channel is bounded as
  \begin{align}
    \snreffopt \geq \frac 1 {4M^2} \sucmin{1}{\Lpt^{-1}}{2}.
  \end{align}
\end{myproposition}
\begin{proof}
  Because $\breve{\mbf P}$ is a unitary matrix of dimension $M$, which is complex-valued, the resulting lattice $\lat{\Lpt^H}$ and its dual are also complex-valued with dimension $M$. In the real-valued domain, those lattices have dimension of $2M$. Hence, following the proof of Proposition~\ref{proposition:snr-bound-orthogonal}, the desired result is obtained.
\end{proof}

From Propositions~\ref{proposition:snr-bound-orthogonal} and \ref{proposition:snr-bound-unitary}, we can see that a larger  $\sucmin{1}{\Lpt^{-1}}{2}$ of the unitary precoding cannot guarantee that the corresponding $\snreffopt$ is also higher than that of the orthogonal precoding. In particular, consider the case of $\sucmin{1}{\Lp^{-1}}{2} \leq \sucmin{1}{\Lpt^{-1}}{2} < 4 \sucmin{1}{\Lp^{-1}}{2} $.\footnote{Note that $\Lp$ corresponds to the orthogonal precoding case, while $\Lpt$ to the unitary case.} Even though  $ \sucmin{1}{\Lpt^{-1}}{2} \geq \sucmin{1}{\Lp^{-1}}{2}$, the corresponding lower bound of $\snreffopt$ of the unitary precoding is lower than that of the orthogonal precoding. Hence, if we search for a precoder matrix over unitary groups, we may end up obtaining lower $\snreffopt$ than in the case when we search over unitary groups even though the optimal unitary precoder matrix found may have larger $\sucmin{1}{\Lpt^{-1}}{2}$. According to \eqref{eq:achievable-rate}  and \eqref{eq:outage-prob}, a lower $\snreffopt$  implies lower achievable rate and higher outage probability. This observation is validated via numerical evaluations presented in Section~\ref{sec:numerical-results}. It is confirmed that indeed even though the average $\sucmin{1}{\Lpt^{-1}}{2}$ is higher than $\sucmin{1}{\Lp^{-1}}{2}$, the orthogonal precoding achieves higher $\snreffopt$ and achievable rate, and lower outage and error probabilities than the unitary precoding. Thus, we can claim that finding the optimal IF-MIMO precoder matrix over orthogonal groups instead of unitary groups is beneficial in terms of both complexity and performance.
% \newpage
\section{Finding the Optimal Precoder Matrix}
\label{sec:find-prec-matr}
To find the optimal orthogonal precoder matrix, let us first define the objective function as follows
\begin{align}
  J(\mbf P) = \min_{\mbf v \in \mbb Z^M \backslash \mbf 0} \norm{\Linv \mbf P \mbf v}^2. \label{eq:objective-function}
\end{align}
The optimization problem in \eqref{eq:main-optim-problem} can now be written as
\begin{align}
  \Popt = \argmax_{\mbf P \in \Om} J (\mbf P). \label{eq:popt-problem}
\end{align}
The difficulties of solving the optimization problem above lie within the combination of two major obstacles: (i) orthogonal matrix constraint and (ii) finding the minimum distance of the lattice $\lat{\Linv \mbf P}$.
 
For a $2 \times 2$ MIMO system, a convenient parameterization of $2$-dimensional orthogonal group was proposed in \cite{SakzadV15UnitaryPrecoding}. The orthogonal matrix $\mbf P$ is parameterized using one angle $\theta$ as
\begin{align}
  \mbf P (\theta) =
  \begin{bmatrix}
    \cos \theta & \sin \theta \\
    -\sin \theta & \cos \theta
  \end{bmatrix}.
\end{align}
With this parameterization, $\Popt$ can be estimated easily by performing a simple exhaustive search over only one parameter $\theta \in [0, \pi/4]$. Indeed, this technique performs very well for $2$-dimensional orthogonal group. However, beyond that, it becomes unwieldy and prohibitively complex because the exhaustive search has to be done over $M(M-1)/2$ parameters (angles) \cite{Raffanetti70} and the minimum distance of the resulting lattice has to be checked at every search or iteration.

A simple approach to solving optimization problems with orthogonality constraint is to perform gradient-based search algorithm such as the steepest gradient (SG) algorithm. Interestingly, by exploiting the geometrical properties of orthogonal group as a Lie group \cite{EdelmanOrtho99,AbrudanEK08,Plumbley05}, the orthogonality constraint is always naturally satisfied at every step of the SG algorithm. This means that an optimization problem with an orthogonality constraint is transformed into an unconstrained one, which makes the optimization process easier. For this reason we will use the SG algorithm on Lie groups \cite{Plumbley05,EdelmanOrtho99,AbrudanEK08} to solve our problem. As general reference for the Lie group theory, see \cite{BCHallLieGroup}. Unfortunately, the SG algorithm on Lie groups is not directly applicable to our problem. This is because our objective function in (\ref{eq:objective-function}) in not purely constrained with orthogonality and it is not even differentiable because it depends not only on $\mbf P$, but also on a discrete integer vector $\mbf v$. To overcome this, we break the problem down into two sub-problems.
\subsection{Sub-Problem 1: Local Search}
\label{sec:sub-problem-1}
Observe that by fixing the integer vector $\mbf v$, we can transform the objective function in (\ref{eq:objective-function}) into a differentiable function on which the SG algorithm can work. Assume that we start the search for the solution from an initial $\Pin \in \Om$. A vector at the minimum distance of $\lat{\Linv \Pin}$ is given by an integer vector $\vin$, i.e., $\sucmin{1}{\Linv \Pin}{} = \norm{\Linv \Pin \vin}$. Define
\begin{align}
  \Jt(\mbf P) = \norm{\Linv \mbf P \vin}^2. \label{eq:Jtilde}
\end{align}
Our first sub-problem is thus, given an initial $\Pin$ with the corresponding $\vin$, find a ``good'' $\Ptopt$ such that
\begin{align}
  \Ptopt = \argmax_{\substack{\mbf P \in \Om,\\ \sucmin{1}{\Linv \mbf P}{} = \norm{\Linv \mbf P \vin}}} \Jt(\mbf P). \label{eq:sub-problem-1}
\end{align}
This means that we must find $\Ptopt$ that maximizes \eqref{eq:Jtilde} such that the minimum distance of $\lat{\Linv \Ptopt}$ is still given by $\vin$, i.e., $\sucmin{1}{\Linv \Ptopt}{} = \normn{\Linv \Ptopt \vin}$. Because $\Om$ is a manifold, we can think geometrically that the search is done by moving over the surface of $\Om$ starting from $\Pin$ to a point that satisfies (\ref{eq:sub-problem-1}). We can also think of this search as rotating the whole lattice points $\Linv \Pin \mbb Z$ until a certain degree such that its minimum distance is maximized while keeping the integer vector giving its minimum distance remains unchanged.

Like the conventional SG algorithm, the search for the solution is done by iteratively moving from one point to another in the search space in the steepest direction. Particularly, at $\ell$-th iteration, a move from the current point $\Pl$ to $\Pln$ over $\Om$ is made. This move is equivalent to the move from $\eye$ to some point $\Rl \in \Om$ such that $\Pln = \Rl\Pl$. The question is then how to choose the movement matrix $\Rl$.

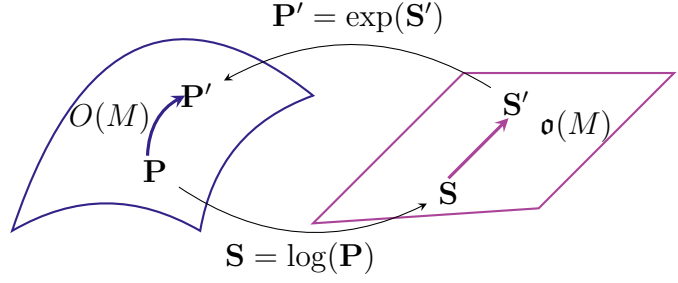
\begin{figure}[t]
  \centering
  \begin{tikzpicture}[thick, >=stealth]
    \draw [TolDarkPurple] (0,0) to [bend left] (2.5,0) to [bend left] (4,1.8) .. controls (2.5,3) and (1,3) .. cycle;
    \draw [->,very thick, TolDarkPurple] (1.8,1) to [bend left] ++(0.5,0.8);
    \draw [TolLightPurple] (4,0.1) -- ++(2,2) -- ++(2.8,0) -- ++(-1.8,-1.8) -- cycle;
    \draw [->, very thick,TolLightPurple] (5.8,0.7) -- ++(0.8,0.8);

    \node[name=P] at  (1.9,0.8) {$\mbf P$};
    \node[name=PPrime] at  (2.45,1.8) {$\mbf P'$};
    \node[name=S] at  (5.8,0.5) {$\mbf S$};
    \node[name=SPrime] at  (6.7,1.7) {$\mbf S'$};

    \draw [->, thin, >=stealth](P) to [bend right] node [below] {$\mbf S = \log (\mbf P)$} (S);
    \draw [->, thin, >=stealth](SPrime) to [bend right] node [above] {$\mbf P' = \exp (\mbf S')$} (PPrime);
    \node at (1.3,1.5) {$\Om$};
    \node at (7.5,1.4) {$\om $};
  \end{tikzpicture}
  \caption{$\Om$ and $\om$ are connected by matrix exponential and logarithm operations \cite{Plumbley05}. A movement over $\Om$ can be defined equivalently by a movement over $\om$.}
  \label{fig:lie-group-algebra}
\end{figure}

For defining a movement in the steepest direction, we will make use of the corresponding Lie algebra $\om$ instead of $\Om$ which is closed only under matrix multiplication. The Lie algebra $\om$ is the vector space of the $M \times M$ skew-symmetric matrices with additional Lie bracket operation in the form of matrix exponential \cite{MolexLoanExpM}. Because $\om$ is a vector space which is closed under addition and scalar multiplication, it is easier to define a movement over $\om$  rather than over $\Om$. $\Om$ and $\om$ are connected by matrix exponential and matrix logarithm operators \cite[Chapter~2]{BCHallLieGroup}. As illustrated in Fig.~\ref{fig:lie-group-algebra}, any point $\mbf P \in \Om$ can be mapped to a point $\mbf S \in \om$ using $\mbf S = \log (\mbf P)$ and any point $\mbf S' \in \om$ can be mapped to a point $\mbf P' \in \Om$ using $\mbf P' = \exp(\mbf S')$. Thus, any movement in $\Om$ is equivalent to a movement in $\om$, and vice versa.

Consider our SG algorithm at  $\ell$-th iteration. To move from $\eye$ to $\Rl$, first, we map $\eye$ to a point in $\om$, which is $\mbf 0$ because $\log (\eye) = \mbf 0$. Then, from $\mbf 0$ we make a move to a point $\Sl$ over $\om$. Once $\Sl$ is found, we can compute $\Rl = \exp (\Sl)$, and subsequently $\Pln = \Rl \Pl$. The movement matrix $\Sl$ has to be decided based on the steepest gradient of $\Jt(\Pl)$ in the $\mbf S$-space. Define $\Delta_{\mbf P} \Jt (\Pl)$ as the gradient of $\Jt (\mbf P)$ in the $\mbf P$-space at $\mbf P = \Pl$. It is easy to derive that
\begin{align}
  \Delta_{\mbf P} \Jt (\Pl) = 2 (\Linv)^2 \Pl \vin \vin^T. \label{eq:delta-P-J}
\end{align}
Using the result from \cite{Plumbley05}, the steepest gradient of $\Jt(\mbf P)$ in the $\mbf S$-space at $\mbf P = \Pl$ is given by
\begin{align}
  \Delta_{\mbf S} \Jt (\Pl) = \Delta_{\mbf P} \Jt (\Pl)\Pl^T - \Pl (\Delta_{\mbf P} \Jt(\Pl))^T. \label{eq:grad-lie-algebra}
\end{align}
For a constant $\mu$, a move from $\mbf 0$ to $\Sl$ now can be defined as
\begin{align}
  \Sl = \mbf 0 + \mu \Delta_{\mbf S} \Jt (\Pl) = \mu \Delta_{\mbf S} \Jt (\Pl). \label{eq:Sl}
\end{align}
We refer to $\mu$ as the \textit{step size}. The move from $\Pl$ to $\Pln$ is thus can be written as
\begin{align}
  \Pln = \exp (\mu \Delta_{\mbf S} \Jt (\Pl)) \Pl.\label{eq:final-move}
\end{align}
 
As in the general SG algorithm, choosing an appropriate step size is crucial for the convergence. A fixed  step size can ensure a convergence close to a local optimum, but in general it requires many iterations. Therefore, it is desirable to select an appropriate step size at each iteration for a faster convergence. The appropriate step size is commonly determined based on the objective function. However, in our problem, the step size depends not only on the objective function, but also on the problem constraint; that is the integer vector providing the minimum distance of the corresponding lattice must not change. To select the appropriate step size at every iteration we propose the following two steps.

\textit{Step 1}: In this step, the step size is determined based on the objective function. Consider a point in $\Om$ emanating from $\Pl$ along the steepest direction $\Delta_{\mbf S} \Jt (\Pl)$ as a function of $\mu$
\begin{align}
  \Pmu = \exp (\mu \Delta_{\mbf S} \Jt (\Pl)) \Pl, \label{eq:pmu}
\end{align}
and define
\begin{align}
  \Jmu \triangleq \Jt(\Pmu). \label{eq:jmu}
\end{align}
The step size at $\ell$-th iteration is chosen such that
\begin{align}
  \mul = \argmax_{\mu} \Jmu. \label{eq:mul}
\end{align}
The optimal $\mul$ is difficult to find in general. Fortunately, our objective function $\Jmu$ in \eqref{eq:mul} has a desirable property that may be exploited to determine $\mul$. The matrix exponential in \eqref{eq:pmu} induces an \textit{almost periodic} \cite{Fischer96,AbrudanEK09} behavior of $\Jmu$ with respect to $\mu$. Thus, to determine $\mul$, we can use existing techniques that are used for finding local minimums of almost periodic functions. In particular, we adopt the polynomial approximation approach proposed in \cite{AbrudanEK09}.

\textit{Step 2}: The $\mul$ obtained in the step 1 is chosen such that $\Jmu$ is maximized. This will not lead us to the solution of (\ref{eq:sub-problem-1}) if the problem constraint is not satisfied, i.e., the integer vector providing the minimum distance of $\lat{\Linv \mbf P(\mul)}$ is different from that of $\lat{\Linv \Pin}$. Therefore the $\mul$ obtained in the step 1 is has to be further adjusted such that the problem constraint is always satisfied. This is easily performed by iteratively halving $\mul$ or dividing $\mul$ by a constant $\zeta > 1$ if the problem constraint in not satisfied.

The summary of the algorithm for solving the sub-problem~1 is presented in Algorithm~\ref{algo:local-search}. To find the minimum distance of a lattice, $\lambda_1(\cdot)$, optimal algorithms such as the Fincke-Pohst \cite{FinckeP85} algorithm or the sphere decoding \cite{ViterboB99SD} algorithm and its variance \cite{Agrell02,SchnorrE94}, may be employed. One can also use the Lenstra-Lenstra-Lov\'asz (LLL) algorithm \cite{Lenstra82LLL} that exhibits much lower complexity. We used the LLL algorithm \cite{LyuBoostedLLL} in our computer simulations.
\begin{algorithm}[t]
  \caption{Local search: Finding a local optimal $\Ptopt$ from an initial orthogonal matrix $\Pin$.}
  \begin{algorithmic}[1]
    \REQUIRE {$\Linv$ and $\Pin$.}
    \ENSURE {An estimate of local optimal precoder matrix $\Ptopt$.}
    \STATE Find $\vin \in \mbb Z^M$ such that $\sucmin{1}{\Linv \Pin}{} = \norm{\Linv \Pin \vin}$.
    \STATE Initialize $\ell=0$, $\Pl = \Pin$.
    \STATE Compute $ \Delta_{\mbf S} \Jt (\Pl)$ as in (\ref{eq:grad-lie-algebra}).
    \STATE Find $\mul$ using the polynomial approximation \cite{AbrudanEK09}.
    \STATE Further adjust $\mul$:\\
    while $\sucmin{1}{\Linv \mbf P (\mul)}{} \neq \norm{\Linv \mbf P (\mul) \vin}$, set $\mul := \mul/2$.
    \STATE Update $\Pln = \mbf P (\mul)$ and $\ell := \ell+1$. Iterate the steps 3~-~6 until convergence or until maximum iteration.
    \RETURN $\Ptopt = \Pl$.
  \end{algorithmic}
  \label{algo:local-search}
\end{algorithm}

\subsection{Sub-Problem 2:  Global Search}
\label{sec:sub-problem-2}
The solution of the sub-problem~1 may not be the global optimal solution because given a starting point $\Pin$, the search is performed over the surface limited to only around $\Pin$. Therefore, to find the global optimal solution, it is crucial to select a good starting point $\Pin$, which becomes our second sub-problem. We state our second sub-problem as follows: from $\Om$, find a good matrix $\Pin$ such that $\sucmin{1}{\Linv \Pin}{}$ is as large as possible. This problem is indeed similar to our original problem in (\ref{eq:popt-problem}), except that the solution of this sub-problem does not have to be optimal. A better or possibly optimal solution will be derived by refining the solution using Algorithm~\ref{algo:local-search}.

To solve this sub-problem, we adopt a random search technique. Random search has been widely used and is very suitable for ill-structured global optimization problem, where the objective function may be not differentiable, and possibly discontinuous over a continuous, discrete, or mixed continuous-discrete domain \cite{Zabinsky09} just like exactly what we have in  (\ref{eq:popt-problem}). Random search in general does not guarantee finding a global optimal solution. But it offers finding a good solution quickly. In literature, it has been shown that random search converge to the global optimal solution with some probability \cite{Zabinsky09,SolisWets81}.

The random search algorithm that we employ is quite straightforward. The algorithm starts by initializing  $\Pin = \eye$. Then, at every iteration $\ell$ an orthogonal matrix $\Pl$ is randomly generated with  Haar measure distribution \cite{Mezzadri07} and the minimum distance of the resulting lattice $\lat{\Linv \Pl}$ is evaluated. If $\sucmin{1}{\Linv \Pl}{} > \sucmin{1}{\Linv \Pin}{}$, then $\Pl$ is kept as the temporary solution, i.e., $\Pin := \Pl$. The more iterations we have, the higher probability that resulting $\Pin$ is close to the global optimal solution $\Popt$. Meanwhile, the complexity also increases. In practice, we do not need many iterations because the result will be further refined using Algorithm~\ref{algo:local-search}. The algorithm for the second sub-problem is summarized in Algorithm~\ref{algo:global-search}.
\begin{algorithm}[t]
  \caption{Global search: Finding a ``good'' initial orthogonal matrix $\Pin$ for Algorithm~\ref{algo:local-search}.}
  \begin{algorithmic}[1]
    \REQUIRE {$\Linv$.}
    \ENSURE {$\Pin \in \Om$ such that $\sucmin{1}{\Linv \Pin}{}$ is large.}
    \STATE Initialize $\ell = 0$, $\Pin := \eye$.
    \STATE Generate a random orthogonal matrix $\Pl$ with Haar measure distribution using \cite{Mezzadri07}.
    \STATE If {$\sucmin{1}{\Linv \Pl}{} > \sucmin{1}{\Linv \Pin}{}$},  $\Pin := \Pl$.
    \STATE $\ell := \ell+1$ and repeat from step 2 for some iterations.
    \RETURN $\Pin$.
  \end{algorithmic}
  \label{algo:global-search}
\end{algorithm}
\subsection{Summary of the Proposed Algorithm}
\label{sec:summ-prop-algo}
To find the solution for our original problem in (\ref{eq:popt-problem}), first, we perform a global search for a good candidate of $\Pin$ over $\Om$ using Algorithm~\ref{algo:global-search}. The resulting $\Pin$ is then used as the starting point of the gradient-based local search following Algorithm~\ref{algo:local-search}, of which the result is expected to be an estimate of the global optimal solution. The overall algorithm is summarized in Algorithm~\ref{algo:overall-search}.
\begin{algorithm}[t]
  \caption{Finding $\Popt$ for the original problem (\ref{eq:main-optim-problem}).}
  \begin{algorithmic}[1]
    \REQUIRE {$\Linv$.}
    \ENSURE {$\Popt$, a solution for (\ref{eq:main-optim-problem}).}
    \STATE Use Algorithm~\ref{algo:global-search} to find $\Pin$.
    \STATE With input $\Pin$, perform Algorithm~\ref{algo:local-search} to obtain $\Ptopt$.
    \STATE Set $\Popt := \Ptopt$.
    \RETURN $\Popt$.
  \end{algorithmic}
  \label{algo:overall-search}
\end{algorithm}

We shall note that the proposed algorithm can also be applied to the unitary precoding case  \cite{SakzadV15UnitaryPrecoding} with some modifications. First, all the regular matrix transpose operations are replaced with the Hermitian transpose. Then, the gradient in (\ref{eq:delta-P-J}) is replaced with $  \Delta_{\mbf P} \Jt (\Pl) = (\Linv)^H \Linv \Pl \vin \vin^H $ and obviously we should generate a random unitary matrix instead of orthogonal one in the step~$2$ of Algorithm~\ref{algo:global-search}. The complexity of the unitary precoding case is clearly higher than the orthogonal precoding because most the operations are done in complex-valued domain rather than real-valued domain.

\section{Discussion of Complexity}
\label{sec:complexity}

\subsection{ Complexity of Algorithm~\ref{algo:overall-search}}
\label{sec:algo-complexity}
This sub-section provides evaluation of computational complexity of Algorithm~\ref{algo:overall-search} and compares it to that of parameterization technique \cite{SakzadV15UnitaryPrecoding}.

%\begin{sloppypar}
The parameterization technique introduced in \cite{SakzadV15UnitaryPrecoding} can be extended to higher dimensional MIMO \cite{Raffanetti70}. In this case, the search for the optimal orthogonal precoder matrix is carried out over at least $M(M-1)/2$ parameters (angles). Denote these parameters as $\theta_1,...,\theta_{M(M-1)/2}$. For simplicity, assume that $\theta_i, \forall i=\{1,..,M(M-1)/2\}$, has a search space of $[0,2\pi)$ which is discretized to $\nu$ samples. For each combination of samples of $\theta_i$, an orthogonal matrix $\mbf P(\theta_1,...,\theta_{M(M-1)/2})$ is constructed and the minimum distance of the resulting lattice $\lat{\Linv \mbf P(\theta_1,...,\theta_{M(M-1)/2})}$  is evaluated. Subsequently $\mbf P(\theta_1,...,\theta_{M(M-1)/2})$ that yields the largest minimum distance of $\lat{\Linv \mbf P(\theta_1,...,\theta_{M(M-1)/2})}$  is chosen as the solution. To keep a low complexity, let us assume that the LLL algorithm with complexity of $\mcal O (M^4\log M)$ \cite{LyuBoostedLLL} is employed. The overall complexity of the parameterization technique is thus $\mcal O(\nu^{M(M-1)/2} M^4 \log M)$.
%\end{sloppypar}

Because the complexity of Algorithm~\ref{algo:overall-search} is dominated by the step 2 where Algorithm~\ref{algo:local-search} is run, we only need to evaluate the complexity of Algorithm~\ref{algo:local-search}. The dominant operations in the Algorithm~\ref{algo:local-search} are finding minimum distance of a lattice in the steps~1 and 5 and calculating matrix exponential in the steps~3, 4, and 5. To find the minimum distance of a lattice, we employ the same LLL algorithm with complexity of  $\mcal O (M^4\log M)$ \cite{LyuBoostedLLL}. While for matrix exponential, there are many ways to calculate it. In literature, we found that the most efficient methods for calculating the matrix exponential exhibit computational complexity of $\mcal O(M^3)$ \cite{MolexLoanExpM}. Because the complexity of finding the minimum distance of a lattice is more dominant, we can ignore the complexity of computing a matrix exponential. Assume that we need $\xi_{\tr i}$ number of iterations to adjust the step size $\mul$ in the step~5 and $\xi_{\tr o}$ number of iterations for Algorithm~\ref{algo:local-search} to converge. Thus, the overall computational complexity of Algorithm~\ref{algo:overall-search} is $\mcal O(\xi_{\tr o} \xi_{\tr i} M^4 \log M)$ or simply $\mcal O(M^4 \log M)$. Now we can clearly see that the complexity of the proposed algorithm is much smaller than that of the parameterization technique.
\subsection{Decoding Complexity}
\label{sec:enc-dec-complexity}
At the receiver side, the decoding complexity of the proposed scheme is nearly the same as ZF and MMSE receivers. This is because the IF receiver manipulates MIMO channels such that a SISO decoding can be employed, which is similar to ZF and MMSE receivers. An additional complexity comes from the step of finding a full-rank integer matrix $\mbf A$. Consider slow-fading channels where the channel coefficients remain constant over a long period called \textit{quasi-static channel interval}. Because $\mbf A$ is essentially an approximation of the MIMO channels which remains constant during the interval, the search for $\mbf A $ needs to be done only once in each static interval. This is in contrast to the general joint ML MIMO decoding in slow-fading channels. Assume that within the static interval, there are $T \in \mbb Z$ number of codeword transmissions that can be made. In the joint ML decoding case, an optimal algorithm such as sphere decoding (SD) algorithm \cite{FinckeP85,ViterboB99SD} which has an exponential complexity has to be performed for each transmission; $T$ times in one static interval. Assume that to find the optimal $\mbf A$, the proposed scheme utilizes the same SD algorithm. In this case, the joint ML decoding would exhibit $T$ times higher complexity than the proposed scheme.

Even though a brute force for finding the optimal integer matrix $\mbf A$ has a high complexity of $\mcal O (\gamma^M)$ \cite{NazerG11}, some effort has been made to develop more efficient algorithms. For instance, Ding \textit{et al.} \cite{DingSMPIF15}  developed an optimal algorithm based on SD and Schnorr-Euchner algorithms \cite{SchnorrE94} to find the optimal $\mbf A$ with computational complexity of $(\pi e) ^{M + \mcal O (\log M)}$. A similar algorithm with a slightly lower complexity was also proposed in \cite{WenIFMIMO17}. To further reduce the complexity, Sakzad \textit{et al.}  \cite{SakzadHV13} proposed an approximation algorithm based on the LLL algorithm with polynomial complexity of $\mcal O(M^4 \log (2M))$. They also investigated algorithms based on Hermite-Korkine-Zolotareff (HKZ) and Minkowski lattice basis reduction algorithms, see \cite{SakzadHV13} for more detail discussion. Other efficient algorithms can be found in \cite{LyuBoostedLLL,Liu16, DingSMPIF15}. 

\section{Numerical Results}
\label{sec:numerical-results}
This section presents and analyzes the numerical results obtained from computer simulations conducted to compare the performance of the proposed schemes with existing schemes. 

\begin{figure}
  \centering
  \begin{subfigure}[t]{0.475\linewidth}
    \centering
    \begin{tikzpicture}
      \begin{axis}[mlineplot,
        scale only axis,
        width = 0.8\linewidth,
        height = 0.8\linewidth,
        xlabel = {SNR $= \gamma$ (dB)},
        ylabel =  Minimum distance,
        legend pos = north west,
        xmin=5, xmax=20, ymin=2,ymax=16,
        xtick=data,
        ]
        \addplot table [x=snrdB, y=dualDMinSGOrtho] {ortho-uni-4x4mimo-rate-mindist.dat};
        \addlegendentry{Orthogonal Prec.}
        \addplot table [x=snrdB, y=dualDMinSGUnit] {ortho-uni-4x4mimo-rate-mindist.dat};
        \addlegendentry{Unitary Prec.}
        \node[name=mdort] at (axis cs:15.8,5.2) {$\mdort$};
        \draw (mdort) -- (axis cs:13.6,7.1);
        \node[name=mduni] at (axis cs:11.1,9.5) {$\mduni$};
        \draw (mduni) -- (axis cs:13.25,7.5);
      \end{axis}
    \end{tikzpicture}
    \caption{Minimun distance of dual lattice}
  \end{subfigure}
  \hfill
  \begin{subfigure}[t]{0.475\linewidth}
    \centering
    \begin{tikzpicture}
      \begin{axis}[mlineplot,
        scale only axis,
        width = 0.8\linewidth,
        height = 0.8\linewidth,
        xlabel = {SNR $= \gamma$ (dB)},
        ylabel =  {Achievable rate},
        legend pos = north west,
        xmin=5, xmax=20, ymin=10, ymax=30,
        xtick=data,
        ]
        \addplot table [x=snrdB, y=rateSGOrtho] {ortho-uni-4x4mimo-rate-mindist.dat};
        \addlegendentry{Orthogonal Prec.}
        \addplot table [x=snrdB, y=rateSGUnit] {ortho-uni-4x4mimo-rate-mindist.dat};
        \addlegendentry{Unitary Prec.}
      \end{axis}
    \end{tikzpicture}
    \caption{Achievable rate}
  \end{subfigure}
  \caption{Performance of the orthogonal and unitary precodings in $4 \times 4$ MIMO : (a) average minimum distance of dual lattices $\lat{\Lp^{-1}}$ (orthogonal) and $\lat{\Lpt^{-1}}$ (unitary), (b) average achievable rate, which is a function of $\snreffopt$ or the largest successive minimum of the corresponding prime lattices.}
  \label{fig:ortho-unitary-dmin-rate}
\end{figure}
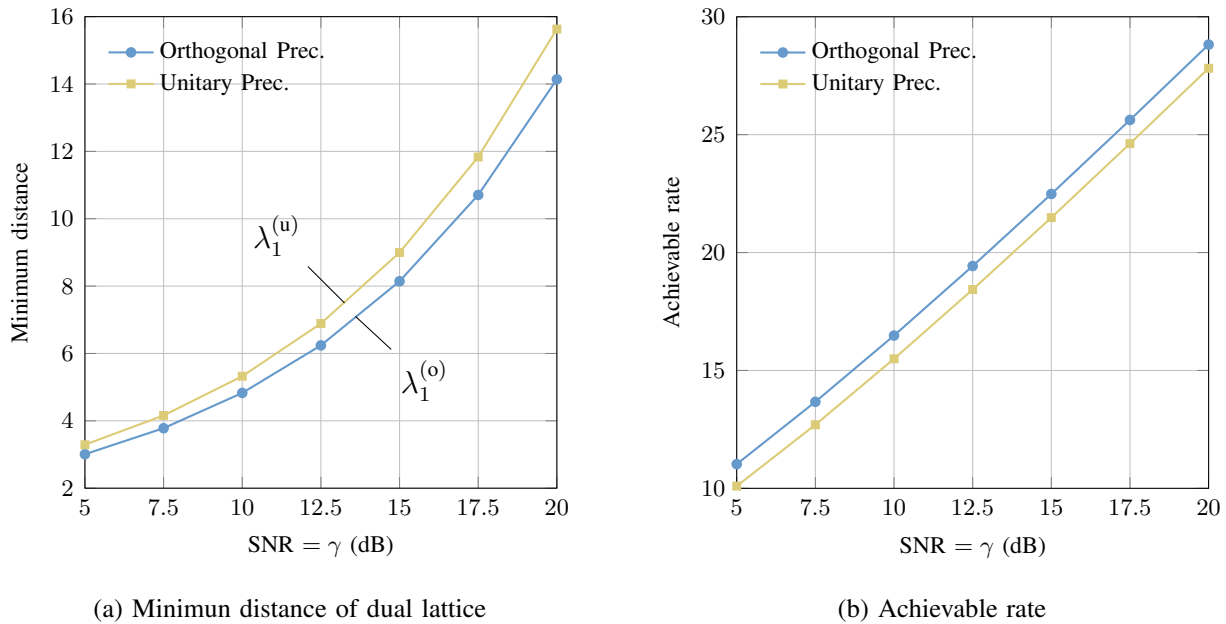
\begin{figure}
  \centering
  \begin{subfigure}[t]{0.475\linewidth}
    \centering
    \begin{tikzpicture}
      \begin{semilogyaxis}[mlineplot,
        scale only axis,
        width = 0.8\linewidth,
        % height=1.4\linewidth,,
        height = 0.8\linewidth,
        xlabel = {SNR $= \gamma$ (dB)},
        ylabel =  Outage Probability,
        legend pos = south west,
        xmin=10, xmax=35, ymin=1e-6,ymax=1,
        ]
        \legend{Orthogonal Prec., Unitary Prec.}
        \addplot[curve1] table [x=SNR_dB, y=ortho_outage] {ortho-uni-4x4mimo-outage-rate24.dat};
        \addplot[curve2] table [x=SNR_dB, y=unit_outage] {ortho-uni-4x4mimo-outage-rate24.dat};
        \addplot[curve1] table [x=SNR_dB, y=ortho_outage] {ortho-uni-4x4mimo-outage-rate32.dat};
        \addplot[curve2] table [x=SNR_dB, y=unit_outage] {ortho-uni-4x4mimo-outage-rate32.dat};

        \draw (axis cs:20,3e-2) ellipse [x radius=1, y radius=.7];
        \draw (axis cs:27.5,1e-2) ellipse [x radius=1, y radius=.7];
        \node at (axis cs:17,3e-2) {\tiny $R_{\tr t} = 24$};
        \node at (axis cs:30,1e-2) {\tiny $R_{\tr t} = 32$};
      \end{semilogyaxis}
    \end{tikzpicture}
    \caption{Outage probability}
  \end{subfigure}
  \hfill
  \begin{subfigure}[t]{0.475\linewidth}
    \centering
    \begin{tikzpicture}
      \begin{semilogyaxis}[mlineplot,
      scale only axis,
      width=0.8\linewidth,
      height=0.8\linewidth,
      xlabel = {SNR $= \gamma$ (dB)},
      ylabel =  {Word-error-rate},
      legend pos = south west,
      xmin=20, xmax=45, ymin=1e-6, ymax=1,
      ]
      \legend{Orthogonal Prec., Unitary Prec.}
      \addplot[curve1] table [x=SNR_dB, y=UPIF_Grad] {wer4x4-64qam.dat};
      \addplot[curve2] table [x=SNR_dB, y=unitary_sg] {wer4x4-64qam-unitary.dat};
      \addplot[curve1] table [x=SNR_dB, y=UPIF_Grad] {wer4x4-256qam.dat};
      \addplot[curve2] table [x=SNR_dB, y=unitary_sg] {wer4x4-256qam-unitary.dat};

      \draw (axis cs:30,1e-2) ellipse [x radius=1, y radius=.7];
      \draw (axis cs:36,1e-2) ellipse [x radius=1, y radius=.7];
      \node at (axis cs:27,1e-2) {\tiny $64$-QAM};
      \node at (axis cs:39,1e-2) {\tiny $256$-QAM};
      \end{semilogyaxis}
    \end{tikzpicture}
    \caption{Word-error-rate}
  \end{subfigure}
  \caption{Performance of the orthogonal and unitary precodings in $4 \times 4$ MIMO: (a) outage probability with target rate $R_{\tr t} \in \{24, 32\}$, (b) word-error-rate with $64/256$-QAM.}
  \label{fig:outage-wer-orthogonal-unitary}
\end{figure}
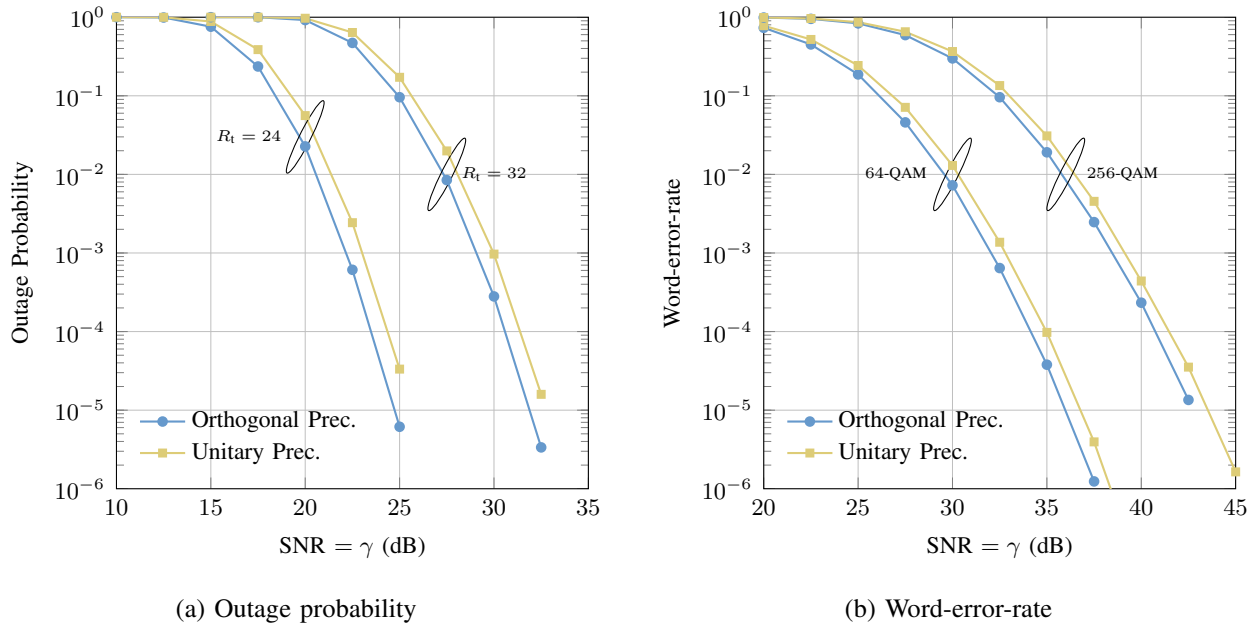

First, we compare the performance of the orthogonal and unitary  precodings.\footnote{Here, the orthogonal and unitary precodings refer to the precoding described in Section~\ref{sec:if-mimo-ortho-precoder} where the precoder matrix is selected from groups of orthogonal and unitary matrices, respectively. The unitary precoding is exactly UPIF~I.} For finding good orthogonal and unitary precoder matrices in the sense of (\ref{eq:main-optim-problem-1}), we use Algorithm~\ref{algo:overall-search} and its modified version described in Subsection~\ref{sec:summ-prop-algo}, respectively. Let $\mdort \triangleq  \sucmin{1}{\mbf \Lp^{-1}}{}$ and $\mduni \triangleq  \sucmin{1}{\mbf \Lpt^{-1}}{}$ denote the minimum distance of the resulting dual lattices of orthogonal and unitary precodings, respectively (cf. Propositions~\ref{proposition:snr-bound-orthogonal} and \ref{proposition:snr-bound-unitary}). Fig.~\ref{fig:ortho-unitary-dmin-rate}(a) shows the average of $\mdort$ and $\mduni$.  Based on Fig.~\ref{fig:ortho-unitary-dmin-rate}(a) and our main optimization problem \eqref{eq:main-optim-problem-1}, one may conclude that unitary precoding is better than the orthogonal precoding because $\mdort$ is larger than  $\mduni$. However, Fig.~\ref{fig:ortho-unitary-dmin-rate}(b) shows the opposite, that orthogonal precoding has higher average achievable rates.
A similar result is shown in Fig.~\ref{fig:outage-wer-orthogonal-unitary} where the orthogonal precoding has lower outage probability and word-error-rate (WER) than unitary precoding.\footnote{We define a word as $({\mbf w}_1, \ldots,{\mbf w}_M)$. For calculating WER, we declare an error event when $(\hat{\mbf w}_1, \ldots,\hat{\mbf w}_M ) \neq ({\mbf w}_1, \ldots,{\mbf w}_M)$.} These results confirm our claim that for the IF-MIMO precoding, in addition to the complexity advantage, searching for precoder matrices over orthogonal groups instead of unitary groups also offers performance advantage. This additional advantage is because the lower bound on $\snreffopt$  of the unitary precoding is smaller than that of the orthogonal precoding as shown in Propositions~\ref{proposition:snr-bound-orthogonal} and \ref{proposition:snr-bound-unitary}. In fact, since the dimension of unitary matrices are twice that of orthogonal matrices in the real-valued domain, the largest successive minimum of the prime lattice $\lat{\Lpt^{H}}$ of the unitary precoding is generally larger than that of the prime lattice $\lat{\Lp^{T}}$ of the orthogonal precoding, and hence its $\snreffopt$ is smaller (see (\ref{eq:snreff-opt})), implying lower achievable rate and higher outage probability. 

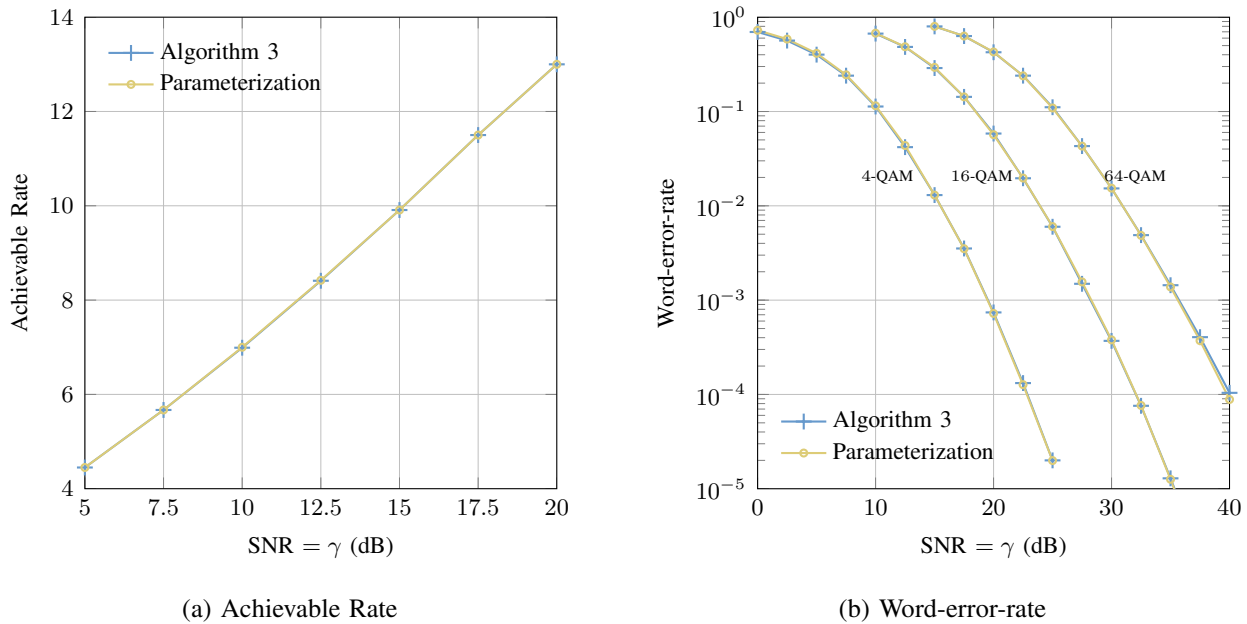
\begin{figure}
  \centering
  \begin{subfigure}[t]{0.475\linewidth}
    \centering
    \begin{tikzpicture}
      \begin{axis}[mlineplot,
        scale only axis,
        width = 0.8\linewidth,
        % height=1.4\linewidth,,
        height = 0.8\linewidth,
        xlabel = {SNR $= \gamma$ (dB)},
        ylabel =  Achievable Rate,
        legend pos = north west,
        xmin=5, xmax=20, ymin=4, ymax=14,
        xtick=data,
        ]
        \legend{Algorithm~\ref{algo:overall-search}, Parameterization}
        \addplot[curve1, mark=+, mark size=3pt] table [ x=snrdB, y= rateT1 ] {mimo2x2Rates.dat};
        \addplot[curve2, mark=o] table [ x=snrdB, y= rateSG ] {mimo2x2Rates.dat};

      \end{axis}
    \end{tikzpicture}
    \caption{Achievable Rate}
  \end{subfigure}
  \hfill
  \begin{subfigure}[t]{0.475\linewidth}
    \centering
    \begin{tikzpicture}
      \begin{semilogyaxis}[mlineplot,
      scale only axis,
      width=0.8\linewidth,
      height=0.8\linewidth,
      xlabel = {SNR $= \gamma$ (dB)},
      ylabel =  {Word-error-rate},
      legend pos = south west,
      xmin=0, xmax=40, ymin=1e-5, ymax=1,
      ]
      \legend{Algorithm~\ref{algo:overall-search}, Parameterization}
      \addplot[curve1, mark=+, mark size=3pt] table [ x=SNR_dB, y= UPIF_T1 ] {wer2x2-4qam.dat};
      \addplot[curve2, mark=o] table [ x=SNR_dB, y= UPIF_Grad ] {wer2x2-4qam.dat};
      \addplot[curve1, mark=+, mark size=3pt] table [ x=SNR_dB, y= UPIF_T1 ] {wer2x2-16qam.dat};
      \addplot[curve2, mark=o] table [ x=SNR_dB, y= UPIF_Grad ] {wer2x2-16qam.dat};
      \addplot[curve1, mark=+, mark size=3pt] table [ x=SNR_dB, y= UPIF_T1 ] {wer2x2-64qam.dat};
      \addplot[curve2, mark=o] table [ x=SNR_dB, y= UPIF_Grad ] {wer2x2-64qam.dat};

      \node at (axis cs:11, 2e-2) {\tiny $4$-QAM};
      \node at (axis cs:19, 2E-02) {\tiny $16$-QAM};
      \node at (axis cs:32, 2E-02) {\tiny $64$-QAM};

      \end{semilogyaxis}
    \end{tikzpicture}
    \caption{Word-error-rate}
  \end{subfigure}
  \caption{Performance of the orthogonal precodings using Algorithm~\ref{algo:overall-search} and parameterization \cite{SakzadV15UnitaryPrecoding} in $2 \times 2$ MIMO: (a) average achievable rate (b) word-error-rate with $4/16/64$-QAM.}
  \label{fig:proposed-vs-parameterization}
\end{figure}

We then compare the performance of the parameterization technique \cite{SakzadV15UnitaryPrecoding} (proposed for UPIF~I) and Algorithm~\ref{algo:overall-search}. The parameterization was proposed in \cite{SakzadV15UnitaryPrecoding} for finding good orthogonal matrices for  $2 \times 2$ IF-MIMO. Even though it is possible to  extend this technique to higher dimension \cite{Raffanetti70}, it exhibits exponential complexity as described in Section~\ref{sec:algo-complexity}. For this reason, we only compared them in the $2 \times 2$ IF-MIMO case.  Fig.~\ref{fig:proposed-vs-parameterization} depicts the results of achievable rate and WER performance of the parameterization algorithm of~\cite{SakzadV15UnitaryPrecoding} (for UPIF-I) compared to our proposed algorithm. It can be clearly seen that Algorithm~\ref{algo:overall-search} achieves nearly identical performance to the parameterization technique in various cases. Since Algorithm~\ref{algo:overall-search} has low complexity and yields good performance, we can easily employ it to realize the orthogonal precoding for higher dimension IF-MIMO as we will see later.

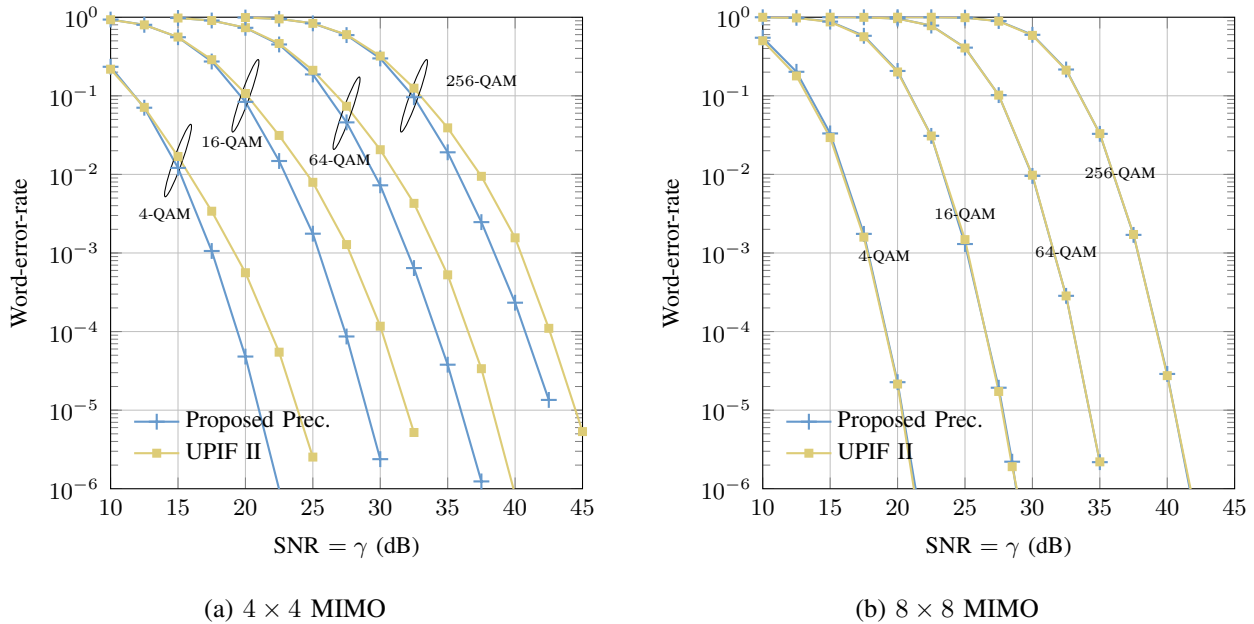
\begin{figure}
  \centering
  \begin{subfigure}[t]{0.475\linewidth}
    \centering
    \begin{tikzpicture}
      \begin{semilogyaxis}[mlineplot,
        scale only axis,
        width = 0.8\linewidth,
        % height=1.4\linewidth,,
        height = 0.8\linewidth,
        xlabel = {SNR $= \gamma$ (dB)},
        ylabel =  Word-error-rate,
        legend pos = south west,
        xmin=10, xmax=45, ymin=1e-6,ymax=1,
        ]
        \legend{Proposed Prec., UPIF~II}
        \addplot[curve1, mark=+, mark size=3pt] table [x=SNR_dB, y=UPIF_Grad] {wer4x4-4qam.dat};
        \addplot[curve2] table [x=SNR_dB, y=UPIF_T2] {wer4x4-4qam.dat};
        \addplot[curve1, mark=+, mark size=3pt] table [x=SNR_dB, y=UPIF_Grad] {wer4x4-16qam.dat};
        \addplot[curve2] table [x=SNR_dB, y=UPIF_T2] {wer4x4-16qam.dat};
        \addplot[curve1, mark=+, mark size=3pt] table [x=SNR_dB, y=UPIF_Grad] {wer4x4-64qam.dat};
        \addplot[curve2] table [x=SNR_dB, y=UPIF_T2] {wer4x4-64qam.dat};
        \addplot[curve1, mark=+, mark size=3pt] table [x=SNR_dB, y=UPIF_Grad] {wer4x4-256qam.dat};
        \addplot[curve2] table [x=SNR_dB, y=UPIF_T2] {wer4x4-256qam.dat};
        
        \draw (axis cs:15,1.5e-2) ellipse [x radius=1, y radius=.7];
        \node at (axis cs:14,3e-3) {\tiny $4$-QAM};
        \draw (axis cs:20,1e-1) ellipse [x radius=1, y radius=.7];
        \node at (axis cs:19,2.5e-2) {\tiny $16$-QAM};
        \draw (axis cs:27.5,6e-2) ellipse [x radius=1, y radius=.7];
        \node at (axis cs:27,1.5e-2) {\tiny $64$-QAM};
        \draw (axis cs:32.5,1e-1) ellipse [x radius=1, y radius=.7];
        \node at (axis cs:37.5,1.5e-1) {\tiny $256$-QAM};

      \end{semilogyaxis}
    \end{tikzpicture}
    \caption{$4 \times 4$ MIMO}
  \end{subfigure}
  \hfill
  \begin{subfigure}[t]{0.475\linewidth}
    \centering
    \begin{tikzpicture}
      \begin{semilogyaxis}[mlineplot,
        scale only axis,
        width=0.8\linewidth,
        height=0.8\linewidth,
        xlabel = {SNR $= \gamma$ (dB)},
        ylabel =  {Word-error-rate},
        legend pos = south west,
        xmin=10, xmax=45, ymin=1e-6, ymax=1,
        ]
        \legend{Proposed Prec., UPIF~II}
        \addplot[curve1, mark=+, mark size=3pt] table [x=SNR_dB, y=UPIF_Grad] {wer8x8-4qam.dat};
        \addplot[curve2] table [x=SNR_dB, y=UPIF_T2] {wer8x8-4qam.dat};
        \addplot[curve1, mark=+, mark size=3pt] table [x=SNR_dB, y=UPIF_Grad] {wer8x8-16qam.dat};
        \addplot[curve2] table [x=SNR_dB, y=UPIF_T2] {wer8x8-16qam.dat};
        \addplot[curve1, mark=+, mark size=3pt] table [x=SNR_dB, y=UPIF_Grad] {wer8x8-64qam.dat};
        \addplot[curve2] table [x=SNR_dB, y=UPIF_T2] {wer8x8-64qam.dat};
        \addplot[curve1, mark=+, mark size=3pt] table [x=SNR_dB, y=UPIF_Grad] {wer8x8-256qam.dat};
        \addplot[curve2] table [x=SNR_dB, y=UPIF_T2] {wer8x8-256qam.dat};

        \node at (axis cs:19,9e-4) {\tiny $4$-QAM};
        \node at (axis cs:25,3e-3) {\tiny $16$-QAM};
        \node at (axis cs:32.5,1e-3) {\tiny $64$-QAM};
        \node at (axis cs:36.5,1e-2) {\tiny $256$-QAM};
        
      \end{semilogyaxis}
    \end{tikzpicture}
    \caption{$8 \times 8$ MIMO}
  \end{subfigure}
  \caption{WER of the proposed precoding and UPIF~II in: (a) $4 \times 4$ MIMO, (b) $8 \times 8$ MIMO.}
  \label{fig:proposed-upif-ii}
\end{figure}

Next, we compare the performance of the proposed orthogonal precoding with UPIF~II. We employ Algorithm~\ref{algo:overall-search} for the proposed precoding. According to \cite{SakzadV15UnitaryPrecoding}, the optimal precoder matrix for UPIF~II should be chosen from \textit{unitary} groups such that it has the largest minimum product distance \cite{FluckigerOV04}. However, finding the minimum product distance of a lattice is a hard problem, especially for unitary matrices. To the best of our knowledge, currently there is no optimal unitary matrix with respect to minimum product distance known. However, there are some available \textit{orthogonal} matrices having good minimum product distance properties listed in \cite{fullDiveRotationsList}. We used these matrices for the UPIF~II simulations. Fig.~\ref{fig:proposed-upif-ii} shows the results of WER for $4 \times 4$ and $8 \times 8$ MIMO configurations each with $4/16/64/256$-QAM. One can see that the proposed precoding and UPIF~II yield nearly the same performance in the $8 \times 8$ MIMO case. While in the $4 \times 4$ MIMO case, the proposed precoding outperforms UPIF~II for all $4/16/64/256$-QAM. Even though we cannot confirm that the proposed precoding is better than UPIF~II for all MIMO configurations, we can say that the proposed precoding can perform better in some scenarios. Moreover, the proposed precoding can be employed for any MIMO dimension, while for dimension beyond $30$, it is hard to realize UPIF~II because no ``good'' orthogonal matrix for UPIF~II with dimension beyond $30$ is currently available in literature. 

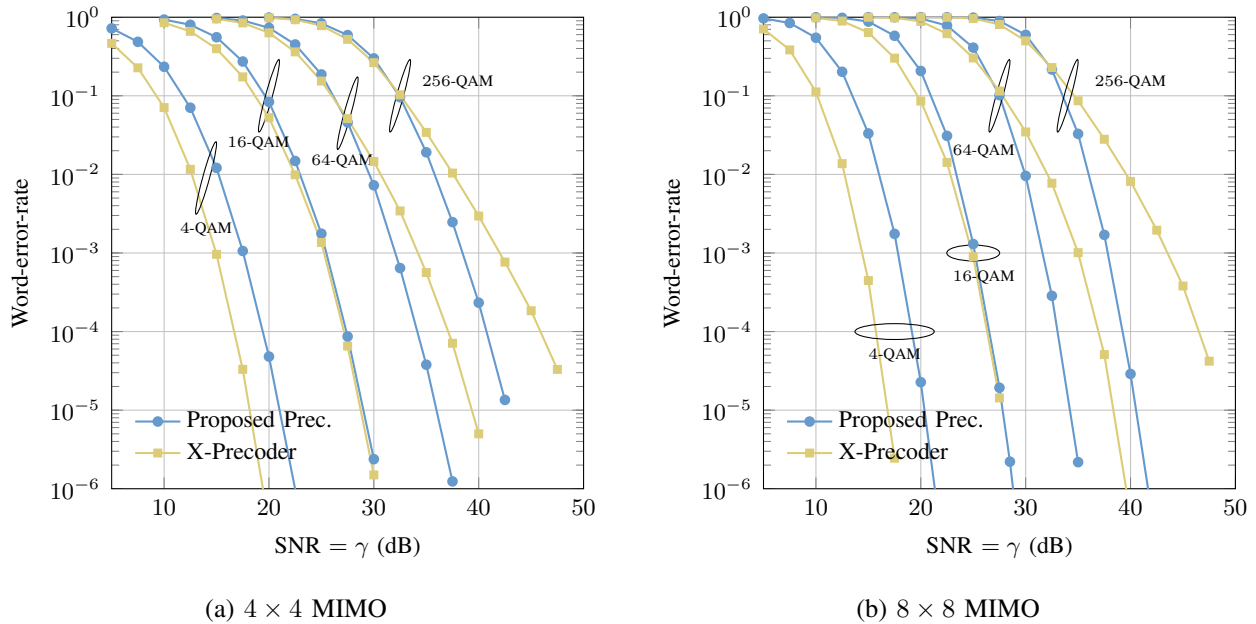
\begin{figure}
  \centering
  \begin{subfigure}[t]{0.475\linewidth}
    \centering
    \begin{tikzpicture}
      \begin{semilogyaxis}[mlineplot,
        scale only axis,
        width = 0.8\linewidth,
        % height=1.4\linewidth,,
        height = 0.8\linewidth,
        xlabel = {SNR $= \gamma$ (dB)},
        ylabel =  Word-error-rate,
        legend pos = south west,
        xmin=5, xmax=50, ymin=1e-6,ymax=1,
        ]
        \legend{Proposed Prec., X-Precoder}
        \addplot[curve1] table [x=SNR_dB, y=UPIF_Grad] {wer4x4-4qam.dat};
        \addplot[curve2] table [x=SNR_dB, y=X_Prec] {wer4x4-4qam.dat};
        \addplot[curve1] table [x=SNR_dB, y=UPIF_Grad] {wer4x4-16qam.dat};
        \addplot[curve2] table [x=SNR_dB, y=X_Prec] {wer4x4-16qam.dat};
        \addplot[curve1] table [x=SNR_dB, y=UPIF_Grad] {wer4x4-64qam.dat};
        \addplot[curve2] table [x=SNR_dB, y=X_Prec] {wer4x4-64qam.dat};
        \addplot[curve1] table [x=SNR_dB, y=UPIF_Grad] {wer4x4-256qam.dat};
        \addplot[curve2] table [x=SNR_dB, y=X_Prec] {wer4x4-256qam.dat};

        \draw (axis cs:14,9e-3) ellipse [x radius=1, y radius=.7];
        \node at (axis cs:14,2e-3) {\tiny $4$-QAM};
        \draw (axis cs:20,1e-1) ellipse [x radius=1, y radius=.7];
        \node at (axis cs:19,2.5e-2) {\tiny $16$-QAM};
        \draw (axis cs:27.5,6e-2) ellipse [x radius=1, y radius=.7];
        \node at (axis cs:27,1.5e-2) {\tiny $64$-QAM};
        \draw (axis cs:32.5,1e-1) ellipse [x radius=1, y radius=.7];
        \node at (axis cs:38,1.5e-1) {\tiny $256$-QAM};
        
      \end{semilogyaxis}
    \end{tikzpicture}
    \caption{$4 \times 4$ MIMO}
  \end{subfigure}
  \hfill
  \begin{subfigure}[t]{0.475\linewidth}
    \centering
    \begin{tikzpicture}
      \begin{semilogyaxis}[mlineplot,
        scale only axis,
        width=0.8\linewidth,
        height=0.8\linewidth,
        xlabel = {SNR $= \gamma$ (dB)},
        ylabel =  {Word-error-rate},
        legend pos = south west,
        xmin=5, xmax=50, ymin=1e-6, ymax=1,
        ]
        \legend{Proposed Prec., X-Precoder}
        \addplot[curve1] table [x=SNR_dB, y=UPIF_Grad] {wer8x8-4qam.dat};
        \addplot[curve2] table [x=SNR_dB, y=X_Prec] {wer8x8-4qam.dat};
        \addplot[curve1] table [x=SNR_dB, y=UPIF_Grad] {wer8x8-16qam.dat};
        \addplot[curve2] table [x=SNR_dB, y=X_Prec] {wer8x8-16qam.dat};
        \addplot[curve1] table [x=SNR_dB, y=UPIF_Grad] {wer8x8-64qam.dat};
        \addplot[curve2] table [x=SNR_dB, y=X_Prec] {wer8x8-64qam.dat};
        \addplot[curve1] table [x=SNR_dB, y=UPIF_Grad] {wer8x8-256qam.dat};
        \addplot[curve2] table [x=SNR_dB, y=X_Prec] {wer8x8-256qam.dat};

        \draw (axis cs:17.5,1e-4) ellipse [x radius=15pt, y radius=3pt];
        \node at (axis cs:17.5,5e-5) {\tiny $4$-QAM};
        \draw (axis cs:25,1e-3) ellipse [x radius=10pt, y radius=3pt];
        \node at (axis cs:26,5e-4) {\tiny $16$-QAM};
        \draw (axis cs:27.5,1e-1) ellipse [x radius=1, y radius=.7];
        \node at (axis cs:26,2e-2) {\tiny $64$-QAM};
        \draw (axis cs:34,1e-1) ellipse [x radius=1, y radius=.7];
        \node at (axis cs:40,1.5e-1) {\tiny $256$-QAM};
        
      \end{semilogyaxis}
    \end{tikzpicture}
    \caption{$8 \times 8$ MIMO}
  \end{subfigure}
  \caption{WER  of the proposed precoding and X-precoder in: (a) $4 \times 4$ MIMO, (b) $8 \times 8$ MIMO.}
  \label{fig:proposed-x-precoder}
\end{figure}

Lastly, we compare the proposed precoding to the X-precoder \cite{XYcodes}, an ML- and QAM-based precoding scheme. In Fig.~\ref{fig:proposed-x-precoder}, we present WER performance for $4 \times 4$ and $8 \times 8$ MIMO configurations with various QAM constellations. In both MIMO configurations, the behavior of WER curves is similar. One can see that X-precoder is better than the proposed precoding for $4$-QAM case, while  for $16$-QAM, both schemes achieve almost the same performance in high SNR regime. However, for $64$- and $256$-QAM, we can clearly see the significant advantage of the proposed precoding over the X-precoder in terms of WER. This advantage comes from the fact that the error performance of the X-precoder is characterized by the minimum distance of received QAM constellations which gets smallers as the constellations size increases. Therefore, the error performace degrades as the constellation size increases. On the other hand, the error performance of the proposed precoding is characterized by the effective SNR, and thus, it is not significantly affected by the constellation size. Moreover, it is known that the X-precoder does not achieve full diversity gain, while similar to UPIF~I \cite{SakzadV15UnitaryPrecoding}, the proposed precoding achieves full diversity gain. We conclude that the proposed orthogonal precoding is superior to the X-precoder for high order QAM.

\section{Conclusions}
\label{sec:conclusions}
We have considered an orthogonal precoding scheme for MIMO with integer-forcing receivers (IF-MIMO). We showed that the proposed orthogonal precoding is better than its unitary counterpart in terms of both performance and complexity. We then proposed methods based on the steepest gradient algorithm on Lie groups and a random search algorithm for finding good orthogonal matrices for the proposed precoding. These methods exhibit lower complexity than the parameterization technique, and can be applied to any MIMO configuration. The numerical results confirmed that the proposed precoding outperforms UPIF~II and the X-precoder in some scenarios. Even though the X-precoder is designed specifically for QAM constellations, the proposed precoding yields better error performance in high order QAM cases, e.g., $64/256$-QAM.

% ========= APPENDICES =========
\appendices

\section{Proof of Proposition~\ref{proposition:snr-bound-orthogonal}}
\label{sec:proof-proposition-ortho}
The proof of Proposition~\ref{proposition:snr-bound-orthogonal} follows the one given in \cite{OrdentlichE15}. Let $\lat{\mbf G}$ be a real-valued lattice generated by a full rank matrix $\mbf G \in \mbb R^{M \times M}$ and let $\lat{\mbf G^{-T}}$ be its dual lattice. In \cite{Banaszczyk93} Banaszczyk proved that the successive minima of $\lat{\mbf G}$ and $\lat{\mbf G^{-T}}$ have the following relationship
\begin{align}
  \sucmin{m}{\mbf G}{}\sucmin{M-m+1}{\mbf G^{-T}}{} \leq M, \label{eq:transference}
\end{align}
for $1 \leq m \leq M$.

From \eqref{eq:snreff-opt}, we have
\begin{align}
  \snreffopt = \frac 1 {\sucmin{M}{\Lp^T}{2}}.
\end{align}
The dual lattice of $\lat{\Lp^T}$ is $\lat{\Lp^{-1}}$, see Definition~\ref{def:dual-lattice}. And thus, by \eqref{eq:transference}, it follows that
\begin{align} 
  \snreffopt \geq \frac {\sucmin{1}{\Lp^{-1}}{2}} {M^2},
\end{align}
which is the desired result.

\section{Proof of Proposition~\ref{proposition:error-prob-bound}}
\label{sec:proof-proposition-error-bound}
Recall that a bijective mapping $\mcal E$ is employed to map $\mbf w_m$ to a codeword $\mbf x_m$. Further, given a full rank matrix $\mbf A$, all $\mbf x_m$'s can be decoded correctly if and only if all sub-channels decode their linear combination $\mbf c_m$ correctly. Therefore, \eqref{eq:error-prob} is equivalent to
\begin{align}
  \Pe &= \Pr\big((\hat{\mbf w}_1,...,\hat{\mbf w}_M) \neq (\mbf w_1,...,\mbf w_M)\big)\\
      &= \Pr\big((\hat{\mbf x}_1,...,\hat{\mbf x}_M) \neq (\mbf x_1,...,\mbf x_M)\big)\\
      &= \Pr\big((\hat{\mbf c}_1,...,\hat{\mbf c}_M) \neq (\mbf c_1,...,\mbf c_M)\big).
\end{align}

Define the error probability at sub-channel $m$ as
\begin{align}
  \label{eq:err-prob-m}
  \Pem = \Pr(\hat{\mbf c}_m \neq \mbf c_m).
\end{align}  
Because $\codlat = \alpha \mbb Z[i]$ and $\shaplat = 2^{2q} \codlat$, the resulting linear combination and effective noise in \eqref{eq:p2p-eff-channel} respectively become $\mbf c_m \in \alpha \mbb Z[i]$ and $\mbf z_{\eff,m} \in \mbb C$, i.e., they are one-dimensional complex-valued vectors. Thus,
\begin{align}
  \Pem &= \Pr\Big(\{\Re(\hat{\mbf c}_m) \neq \Re(\mbf c_m)\} \cup \{\Im (\hat{\mbf c}_m) \neq \Im (\mbf c_m)\}\Big) \nonumber\\
       &\leq 2\Pr\big(\Re(\hat{\mbf c}_m) \neq \Re(\mbf c_m)\big) \label{eq:err-prob-m-re-im}\\
       &= 2 \Pr \Big(\abs{\Re(\mbf z_{\eff,m})} \geq \frac \alpha 2 \Big)\label{eq:slice-error}\\
       &= 4 \Pr \Big(\Re(\mbf z_{\eff,m}) \geq \frac \alpha 2 \Big)\label{eq:slice-error-2},
\end{align}
where \eqref{eq:err-prob-m-re-im} is due to union bound and the fact that $\Re(\mbf c_m)$ and $\Im(\mbf c_m)$ have an identical probability distribution, \eqref{eq:slice-error} is because $\Re(\mbf c_m)$ and $\Im(\mbf c_m)$ are decoded using the nearest-neighbor quantizer with respect to $\alpha \mbb Z[i]$, and \eqref{eq:slice-error-2} follows the symmetry of probability density function of $\Re(\mbf z_{\eff,m})$ around zero. Using \cite[Lemma 4]{OrdentlichE15}, we have
\begin{align}
  \Pem &\leq 4 \Pr \Big(\Re(\mbf z_{\eff,m}) \geq \frac \alpha 2 \Big) \\
       &\leq 4 \exp\left( -\frac {\alpha^2} {4\sigma_{\eff,m}^2}\right)\\
       &= 4 \exp\left( -\frac {\alpha^2} {4\gamma\norm{\Lp^T \mbf a_m}^2}\right).
\end{align}
If $\mbf A_\opt$ is employed, then
\begin{align}
  \Pem &\leq 4 \exp\left( -\frac {\alpha^2} {4\gamma\sucmin{m}{\Lp^{T}}{2}}\right)\\
       &= 4 \exp\left( -\frac {3} {2^{4q+1}\sucmin{m}{\Lp^{T}}{2}}\right)
\end{align}
Now, due to \eqref{eq:nondec-suc-min}, for all $m = \{1,...,M\}$, we have
\begin{align}
  \Pem  &\leq 4 \exp\left( -\frac {3} {2^{4q+1}\sucmin{M}{\Lp^{T}}{2}}\right)\\
        &\leq 4 \exp\left( -\frac {3\sucmin{1}{\Lp^{-1}}{2}} {2^{4q+1}M^2}\right) \label{eq:pem-bound},
\end{align}
where \eqref{eq:pem-bound} follows \eqref{eq:transference}.

With union bound, we derive the total error probability of the system as
\begin{align}
  \Pe       &= \Pr\big((\hat{\mbf c}_1,...,\hat{\mbf c}_M) \neq (\mbf c_1,...,\mbf c_M)\big)\\
            &\leq \sum_{m=1}^M \Pem\\
            &= 4M \exp\left( -\frac {3\sucmin{1}{\Lp^{-1}}{2}} {2^{4q+1}M^2}\right),
\end{align}
which completes the proof.
% ========= BIBLIOGRAPHY =========
\bibliographystyle{IEEEtran}
\bibliography{HasanBibtex}
\end{document}